\documentclass[journal]{IEEEtran}
\usepackage{bm}
\usepackage{cite}
\usepackage{amsmath}
\usepackage{amsthm}
\usepackage{amssymb}
\usepackage{xcolor}
\usepackage{enumitem}
\usepackage[utf8]{inputenc}
\usepackage{graphicx}
\usepackage{float}
\usepackage[colorlinks,linkcolor=black,urlcolor=black,anchorcolor=black,citecolor=black,hyperfootnotes=true]{hyperref}
\usepackage{color}
\usepackage{subfigure}
\usepackage{diagbox}
\usepackage{verbatim}
\usepackage{algorithm}
\usepackage{algorithmic}
\usepackage{amsmath,epsfig,amssymb,subfigure,bm,dsfont}
\usepackage{lipsum}
\newtheorem{theorem}{Theorem}
\newtheorem{lemma}{Lemma}

\newtheorem{remark}{Remark}
\theoremstyle{definition}
\newtheorem{example}{Example}
\newtheorem{definition}{Definition}

\title{{Device-Free Sensing in OFDM Cellular Network}
\thanks{Manuscript received August 20, 2021; revised December 7, 2021; accepted January 14, 2022. This work was supported in part by the Research Grants Council, Hong Kong, China, under Grant 25215020; in part by The Hong Kong Polytechnic University Start-Up Fund, under Grant P0036248; in part by the National Key R\&D Program of China with grant No. 2018YFB1800800; in part by the Basic Research Project No. HZQB-KCZYZ-2021067 of Hetao Shenzhen-HK S\&T Cooperation Zone; in part by Shenzhen Outstanding Talents Training Fund 202002; in part by Guangdong Research Projects No. 2017ZT07X152 and No. 2019CX01X104. {\it (Corresponding author: Liang Liu.)}}
\thanks{Q. Shi, L. Liu, and S. Zhang are with the Department of Electronic and Information Engineering, The Hong Kong Polytechnic University, Hong Kong SAR, China (e-mails: qin-eie.shi@connect.polyu.hk, \{liang-eie.liu,shuowen.zhang\}@polyu.edu.hk).}
\thanks{S. Cui is with the School of Science and Engineering (SSE) and Future Network of Intelligence Institute (FNii), the Chinese University of Hong Kong, and Shenzhen Research Institute of Big Data, Shenzhen, China, 518172; he is also affiliated with Peng Cheng Laboratory, Shenzhen, China, 518066 (e-mail: shuguangcui@cuhk.edu.cn).}
}
\author{\IEEEauthorblockN{Qin Shi, Liang Liu, Shuowen Zhang, and Shuguang Cui}}

\begin{document}

\maketitle \thispagestyle{empty} \vspace{-0.3in}

\begin{abstract}
This paper considers device-free sensing in an orthogonal frequency division multiplexing (OFDM) cellular network to enable integrated sensing and communication (ISAC). A novel two-phase sensing framework is proposed to localize the passive targets that cannot transmit/receive reference signals to/from the base stations (BSs), where the ranges of the targets are estimated based on their reflected OFDM signals to the BSs in Phase I, and the location of each target is estimated based on its ranges to different BSs in Phase II. Specifically, in Phase I, we design a model-free range estimation approach by leveraging the OFDM channel estimation technique for determining the delay values of all the two-way BS-target-BS paths, which does not rely on any BS-target channel model. In Phase II, we reveal that ghost targets may be falsely detected in some cases as all the targets reflect the same signals to the BSs, which thus do not know how to match each estimated range with the right target. Interestingly, we show that the above data association issue is not a fundamental limitation for device-free sensing: under the ideal case of perfect range estimation in Phase I, the probability for ghost targets to exist is proved to be negligible when the targets are randomly located. Moreover, under the practical case of imperfect range estimation in Phase I, we propose an efficient algorithm for joint data association and target localization in Phase II. Numerical results show that our proposed two-phase framework can achieve very high accuracy in the localization of passive targets, which increases with the system bandwidth.
\end{abstract}


\IEEEpeerreviewmaketitle

\section{Introduction}
\subsection{Motivation}
Radar and wireless communication are probably the two most successful applications of radio technology over the past decades. Recently, there has been growing interests in achieving integrated sensing and communication (ISAC) under a common system via reusing the same radio frequency (RF) signals due to its significant benefits brought to many use cases \cite{sturm2011waveform,paul2016survey,zheng2019radar,Vorobyov19,Hassanien19,liu2020joint,Tan21,Zhang21}. For example, the intelligent transportation system can take advantage of the ISAC techniques for sensing the environment and disseminating the sensed data among vehicles to improve the traffic efficiency and safety. Moreover, ISAC techniques can play a crucial role in future communication systems as well. For instance, sensing information in millimeter wave (mmWave) systems can be leveraged to design efficient beam selection and alignment \cite{muns2019beam}.

Despite the appealing future promised by ISAC techniques, how to realize the functions of sensing and communication simultaneously in a practical system is still an open problem. Motivated by this, in this paper, we devote our endeavor to the study of advanced signal processing techniques for estimating the locations of targets by leveraging the communication signals sent by the base stations (BSs) in the cellular network. The ultimate goal of this line of research is to pave the way for transforming the cellular network into a huge sensor, such that the new function of networked localization can be provided to the users in the future beyond-fifth-generation (B5G) and sixth-generation (6G) cellular networks.

\subsection{Prior Work}
The study of ISAC techniques is still in its infancy. However, there are many interesting and important explorations made recently for this emerging direction, as discussed in the following.
\subsubsection{Radar Signal Based Versus Communication Signal Based ISAC}
Intuitively, we can use either the radar signals or the communication signals to achieve ISAC. For the former direction, the key challenge lies in how to embed information into the radar signals; while for the latter direction, the key challenge lies in how to localize the targets using the communication signals. Although several interesting works have been done to modulate a small number of bits into the radar signals \cite{hassanien2015dual,Sahin17}, this approach cannot achieve high-rate data transmission that is necessary for many ISAC applications (e.g., autonomous cars may generate a huge amount of sensing data to be exchanged among adjacent cars in a short time), since modulating high-order random data symbols on the radar signals will significantly reduce the autocorrelation between the transmitted and reflected signals, thus deteriorating the sensing performance. Motivated by this limitation, this paper aims to exploit the use of communication signals in the cellular network for achieving target localization with an accuracy level similar to that achieved by the radar system.
\subsubsection{Device-Based Versus Device-Free ISAC}
Along the line of communication signal based ISAC, the sensing techniques can be further divided into two categories: device-based sensing for localizing registered targets with communication capabilities, and device-free sensing for localizing unregistered targets that cannot transmit/receive communication signals.

Device-based sensing estimates the target locations based on a set of wireless reference signals exchanged between the targets and the BSs \cite{Anderson07,Radar17}, and has been available in the cellular network since the second generation (2G), e.g., the location of a mobile phone can be estimated when it makes an emergency call. Typical methods include time-of-arrival (ToA) based localization which estimates each target's location at the intersection of at least three circles whose radii are products of the speed of the light and the signal propagation time, and angle-of-arrival (AoA) based localization which estimates each target's location at the intersection of lines formed by measuring the arrival angles of radio signals between the target and multiple BSs.

On the other hand, for passive targets that do not have communication capabilities or are unregistered in the network, device-free sensing needs to be leveraged for their localization. Note that in device-free sensing, the cellular network can only estimate the locations of the targets based on their \emph{reflected} communication signals (instead of the actively exchanged reference signals in device-based sensing), similar to the radar systems. However, the signal processing techniques used in radar systems cannot be applied because the communication signals usually do not have an ambiguity function with steep and narrow main lobes. This thus motivates our work on developing new methods for device-free sensing.

\begin{figure}
\centering
\includegraphics[scale=0.39]{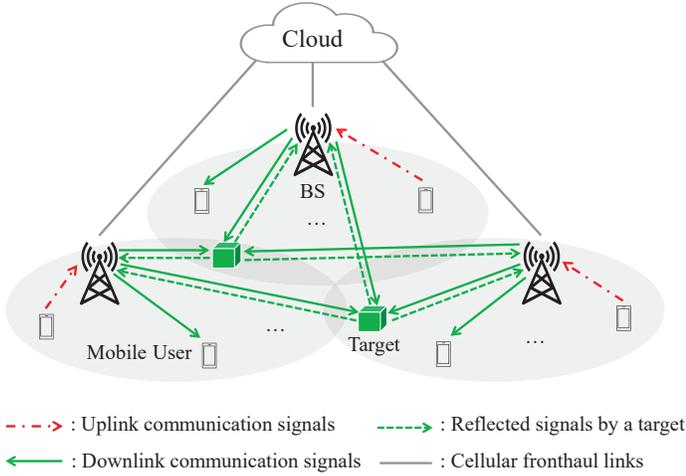}
\caption{System model for an ISAC network with simultaneous downlink/uplink communication and target sensing.}
\label{fig1}
\end{figure}

\subsubsection{Fundamental Limits Versus Practical Solutions}
For device-free sensing based on cellular communication signals, there are generally two research directions: revealing fundamental limits on the communication-sensing trade-off, and designing practical sensing solutions.

Firstly, the optimal waveform design for communication is generally different from that for sensing due to the distinct objectives (i.e., to maximize the mutual information versus to minimize the sensing error), thus leading to a fundamental trade-off between the capacity in communication and the estimation distortion in sensing. Several pioneering works have been done to characterize such an important capacity-distortion trade-off, to reveal the performance upper bound of ISAC systems (see, e.g., \cite{An21,Caire19,Zhang11,Chiriyath16,Masouros18,Masouros18radar}). Secondly, it is also crucial to design practical signal processing solutions for approaching the above fundamental limits. Note that the BS-target-BS reflected channel is generally a function of the target location, thus making it possible to extract the location information by exploiting the reflected channel that can be estimated via channel training. To achieve this goal, several prior works have proposed advanced algorithms based on knowledge of the reflected channel models \cite{kumari2017ieee,Rahman20,Shahmansoori18,Buzzi19,liu2020}. However, the exact BS-target-BS reflected channel models are generally difficult to obtain in practice due to the complicated and time-varying wireless environment, while inexact channel models that do not match with the actual channels will lead to erroneously estimated location even if channel estimation is perfect. This thus motivates us to propose a \emph{model-free} scheme for localizing passive devices that does not depend on knowledge of the reflected channel model, similar to device-based sensing.

\subsection{Main Contributions}
In this paper, we aim to devise practical solutions to achieve device-free sensing in a cellular network, where multiple BSs and multiple mobile users send downlink and uplink communication signals, respectively, while the BSs also collaboratively estimate the locations of multiple passive targets based on the downlink communication signals reflected by the targets, as illustrated in Fig. \ref{fig1}. In particular, we consider the \emph{orthogonal frequency division multiplexing (OFDM)} scheme for communication signal transmission, thus our results are compatible with the 5G and beyond cellular networks. Under this setup, we propose advanced signal processing techniques for localizing the passive targets based on their reflected OFDM signals back to the BSs. The main contributions of this paper are summarized as follows.
\begin{itemize}[leftmargin=*]
\item First, we propose a novel \emph{two-phase framework for device-free sensing} as shown in Fig. \ref{fig12}. Specifically, in Phase I, each BS estimates the values of its distance (also termed as \emph{range}) to the multiple targets by extracting the delay information embedded in the BS-target-BS channels; then, in Phase II, all the BSs share their range information through the fronthaul links in the cellular network (as illustrated in Fig. \ref{fig1}), such that the location of each target can be estimated based on the values of its distance to different BSs, similar to the ToA-based localization approach \cite{Anderson07,Radar17}.

\begin{figure}
\centering
\includegraphics[scale=0.43]{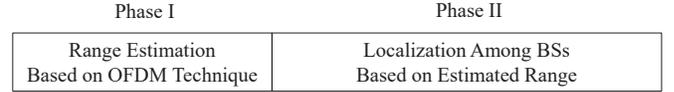}
\caption{Proposed two-phase framework for device-free sensing in OFDM-based ISAC cellular networks.}
\label{fig12}
\end{figure}

\item Second, for Phase I in our considered framework, we propose a new \emph{model-free} scheme to estimate the values of the distance between the BSs and the targets. Specifically, we identify that the signal propagation from the BSs to the targets and then back to the BSs automatically forms a \emph{multi-path channel}, where each BS-target-BS link can be viewed as a (delayed) path between the transmitter and the receiver. Such channels are similar to those in wideband communications, thus the channels at different delayed taps can be efficiently estimated using mature techniques in OFDM communications \cite{Li09}. Note that if the channel tap associated with a particular delay value is non-zero, a path causing that delay exists between the transmitter and the receiver. Inspired by the above, we propose to first estimate the non-zero channel taps and their associated delay values between the BSs and the targets, and then estimate the range of each target as half of the reflection delay multiplied by the speed of the light. The range estimation accuracy is shown to increase with the channel bandwidth, thus is practically high due to the sufficiently large bandwidth in future cellular networks (e.g., up to $400$ MHz in the 5G network \cite{3gpp}). Note that in contrast to the conventional device-free sensing approaches \cite{kumari2017ieee,Rahman20,Shahmansoori18,Buzzi19}, our proposed range estimation method does not depend on the assumption of any BS-target channel model for extracting the distance information.

\item Third, note that the target localization in Phase II of the proposed framework has a key difference from the conventional ToA-based localization. Specifically, in ToA-based localization, different active targets can transmit/receive signals with different signatures such that each BS has a clear mapping between different ranges and different targets. On the contrary, in device-free sensing, all the targets reflect the same signals to the BSs, thus the BSs do not directly know how to match the ranges with the right targets. In the literature, such a matching process is referred to as \emph{data association} \cite{Mahler07}, and it is well-known that an incorrect data association solution may result in ghost targets that do not exist \cite{Mahler07,aditya2017localization}. To tackle this challenge, we first consider the ideal case with perfect range estimation in Phase I, and prove that ghost targets never exist when the number of BSs is more than twice of the number of targets, and do not exist almost surely even when the number of BSs is much smaller than the number of targets. As a result, the ghost target issue arising from data association is not a fundamental limitation for device-free sensing. Moreover, in the case with imperfect range estimation in Phase I, we propose a maximum-likelihood (ML) based algorithm to match each range with the right target, and then estimate the location of each target based on its matched ranges to different BSs.
\end{itemize}
\subsection{Organization}
The rest of this paper is organized as follows. Section \ref{sec:System Model} describes the ISAC system model. Section \ref{sec:Two-Phase Target Sensing Framework} introduces the proposed framework for OFDM-based device-free sensing. Section \ref{sec:Range Estimation} presents the method to estimate the values of distance between the BSs and the targets. Section \ref{sec:Localization with Perfect Range Estimation} studies the fundamental limits for the ghost target existence probability in the ideal case with perfect range estimation; and Section \ref{sec:Localization with Imperfect Range Estimation} proposes an ML algorithm for joint data association and localization in the practical case with imperfect range estimation. At last, Section \ref{sec:Concluding Remarks} concludes this paper and points out some interesting future research directions.

\section{System Model}\label{sec:System Model}
\subsection{Device-Free Sensing Network}
In this paper, we consider an OFDM-based ISAC cellular system that consists of $M\geq 3$ BSs,\footnote{We consider $M\geq 3$ since at least $3$ BSs are needed for localization even in device-based sensing where targets can transmit/receive reference signals.} denoted by $\mathcal{M}=\{ 1,\ldots,M \}$; $I\geq 1$ mobile users for communication, denoted by $\mathcal{I}=\{ 1,\ldots,I \}$; and $K\geq 1$ targets without communication capability for localization, denoted by $\mathcal{K}=\{ 1,\ldots,K \}$, as illustrated in Fig. \ref{fig1}. Besides, we assume that adjacent BSs use orthogonal frequencies \cite{5gstandard,Saquib13,Liu19}, therefore the interference from remote BSs that reuse the same sub-carriers are negligible. Under a two-dimensional (2D) Cartesian coordinate system, the locations of the $k$-th target and the $m$-th BS are denoted as $(x_k,y_k)$ and $(a_m,b_m)$ in meter (m), respectively, $\forall k \in \mathcal{K}$ and $\forall m \in \mathcal{M}$. Thus, the distance between the $m$-th BS and the $k$-th target is given by $d_{m,k}=\sqrt{(a_m-x_k)^2+(b_m-y_k)^2}$ m, and that between the $m$-th BS and the $u$-th BS is given by $d_{m,u}^{{\rm BS}}=\sqrt{(a_m-a_u)^2+(b_m-b_u)^2}$ m, $\forall k\in \mathcal{K}$ and $\forall m,u\in \mathcal{M}$.\footnote{Similar to \cite{Anderson07,Radar17,aditya2017localization}, this paper focuses on 2D localization for the purpose of exposition. However, our results can also be extended to three-dimensional (3D) localization by considering \hbox{an additional height coordinate for each target.}} In the above ISAC system, the BSs send downlink communication signals to the mobile users, while the mobile users send uplink communication signals to the BSs. Moreover, the downlink communication signals from the BSs are reflected by the targets back to the BSs, based on which the cellular network can estimate the locations of the targets as well, similar to radar systems.

\begin{figure}
\centering
\includegraphics[scale=0.25]{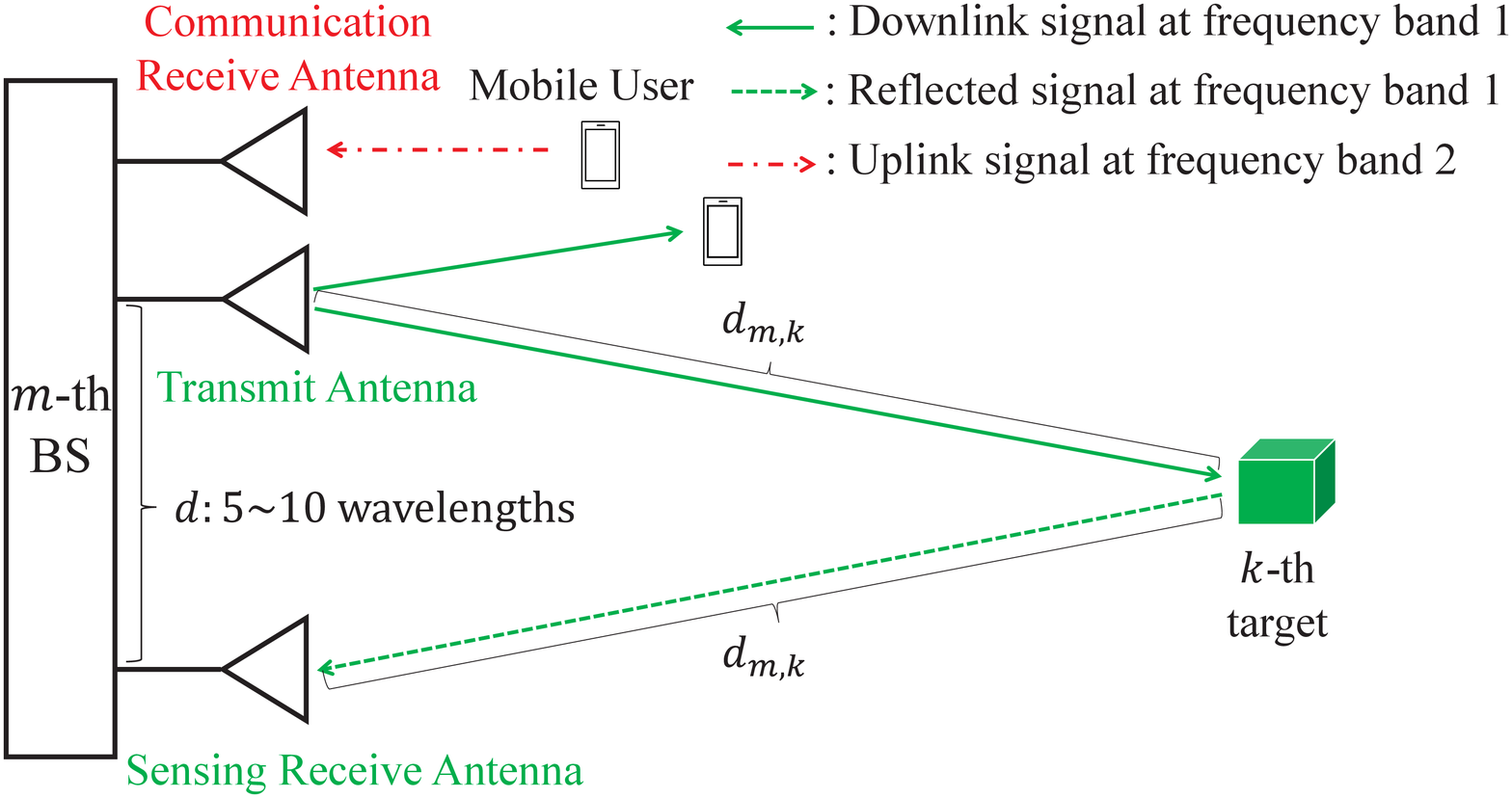}
\caption{Proposed FDD architecture at the BSs for ISAC.}
\label{fig2}
\end{figure}

To simultaneously enable downlink communication, uplink communication, as well as target sensing, this paper proposes a \emph{frequency-division duplexing (FDD)} based ISAC framework. Specifically, as shown in Fig. \ref{fig2}, each BS is equipped with one transmit antenna working at frequency band 1 for transmitting the downlink communication signals to the mobile users, one communication receive antenna working at frequency band 2 that is non-overlapping with frequency band 1 for receiving the uplink communication signals from the mobile users, and one sensing receive antenna working at frequency band 1 to receive the downlink signals reflected by the targets. Under this FDD architecture, the communication receive antennas will not receive interference from the signals reflected by the targets, while the sensing receive antennas will not receive interference from the uplink signals sent by the mobile users. However, each sensing receive antenna will receive strong self-interference from the transmit antenna at the same BS, since each BS works in the full-duplexing mode at frequency band 1. To deal with this issue, we propose to use the techniques of RF isolation and digital-domain cancellation together to mitigate the self-interference \cite{everett2014passive}. Specifically, RF isolation works well in practice when the distance between the transmit and receive antennas is equal to $5-10$ wavelengths of the RF signals. For example, considering a typical 5G carrier frequency at $3.5$ GHz, the transmit antenna and sensing receive antenna can be placed around $0.43-0.86$ m away from each other at each BS to achieve RF isolation. Then, since each BS knows its transmit signals, the self-interference can be cancelled in the the digital domain efficiently. Note that in practice, the distance between a remote target and the BS is much larger than the aforementioned RF isolation distance between antennas. As a result, we assume in this paper that the distance between target $k$ and the transmit antenna at BS $m$ as well as that between target $k$ and the sensing receive antenna at BS $m$ are both equal to $d_{m,k}$, $\forall k\in \mathcal{K}$ and $\forall m\in \mathcal{M}$, as illustrated in Fig. \ref{fig2}.

\subsection{Sensing Signal Model}
Since the OFDM cellular communication technology is quite mature, in the rest of this paper, we mainly study how to leverage the OFDM communication signals for sensing the targets in our considered ISAC cellular system. Let $N$ and $\Delta f$ (in Hz) denote the number of sub-carriers and the sub-carrier spacing of the downlink OFDM signals, respectively, thus the channel bandwidth is $B=N \Delta f$ Hz. Then, in the baseband domain, define
\begin{equation}
\boldsymbol{\chi}_m = [\chi_{m,1},\ldots, \chi_{m,N}]^T = \sqrt{p}\boldsymbol{W}^H \boldsymbol{s}_m, ~~~ \forall m,
\label{eqS2.3}
\end{equation}
as a time-domain OFDM symbol transmitted from BS $m$ consisting of $N$ OFDM samples, $\chi_{m,1},\ldots, \chi_{m,N}$, where $p$ denotes the common transmit power at the BSs; $\boldsymbol{s}_m = [s_{m,1},\ldots,s_{m,N}]^T$ denotes the frequency-domain OFDM symbol with $s_{m,n}$ denoting the unit-power signal at the $m$-th BS over the $n$-th sub-carrier; and $\boldsymbol{W} \in \mathbb{C}^{N \times N}$ denotes the discrete Fourier transform (DFT) matrix with $\boldsymbol{W}\boldsymbol{W}^H=\boldsymbol{W}^H\boldsymbol{W}=\boldsymbol{I}$. Note that the lengths of each OFDM symbol period and each OFDM sample period are $1/\Delta f$ seconds (s) and $1/(N \Delta f)$ s, respectively.

Before the beginning of each OFDM symbol, a cyclic prefix (CP) consisting of $Q < N$ OFDM samples is inserted to eliminate the inter-symbol interference. The overall time-domain transmitted signal by the $m$-th BS for one OFDM symbol is thus expressed as
\begin{align}
\bar{\boldsymbol{\chi}}_m   = & [\underbrace{\bar{\chi}_{m,-Q},\ldots,\bar{\chi}_{m,-1}}_{\text{CP}},\underbrace{\bar{\chi}_{m,0},\ldots,\bar{\chi}_{m,N-1}}_{\text{pilot or data}}]^T\nonumber \\ = & [\underbrace{\chi_{m,N-Q+1},\ldots,\chi_{m,N}}_{\text{CP}},\underbrace{\chi_{m,1},\ldots,\chi_{m,N}}_{\text{pilot or data}}]^T, ~ \forall m.
  \label{eqS2,5}
\end{align}

In this paper, we neglect the signals that are reflected by more than one target since they are generally too weak to be detected at the sensing receive antennas. The received signal at the sensing receive antenna at each BS is thus the superposition of the receiver noise and the downlink OFDM signals sent from all the $M$ BSs, each reflected by all the $K$ targets. Note that this automatically constitutes a \emph{multi-path channel} between the BSs' transmit antennas and the BSs' sensing receive antennas, with each target serving as a scatter that causes a delayed path. Let $L$ denote the maximum number of resolvable paths, with $L<Q$. The received signal at the sensing receive antenna at the $m$-th BS in the $n$-th OFDM sample period is thus expressed as
\begin{equation}
	y_{m,n}=\sum_{u=1}^M \sum_{l=1}^L h_{u,m,l} \bar{\chi}_{u,n-l}+z_{m,n},\quad \forall m,n,
	\label{eqS2.6}
\end{equation}
where $h_{u,m,l}$ denotes the complex channel for the path from the $u$-th BS to the $m$-th BS scattered by a target that causes a delay of $l$ OFDM sample periods, and $z_{m,n} \sim \mathcal {CN}(0,\sigma_z^2)$ denotes the circularly symmetric complex Gaussian (CSCG) noise at the sensing receive antenna of the $m$-th BS during the $n$-th OFDM sample period, with $\sigma_z^2$ denoting the average noise power.

After removing the first $Q$ samples corrupted by the CP, the received signal at the sensing receive antenna of the $m$-th BS over one OFDM symbol period can be expressed as
\begin{equation}
\boldsymbol{y}_m=\sum_{u=1}^M \boldsymbol{H}_{u,m} \boldsymbol{\chi}_u+\boldsymbol{z}_m,\quad \forall m,
\label{eqS2.7}
\end{equation}
where $\boldsymbol{y}_m=[y_{m,1},\ldots,y_{m,N}]^T$; $\boldsymbol{H}_{u,m}$ is an $N\times N$ circulant matrix with the first row defined as $[h_{u,m,1}, 0, \cdots, 0,  h_{u,m,L}, \cdots, h_{u,m,2}]\in \mathbb{C}^{1\times N}$; and $\boldsymbol{z}_m=[z_{m,1},\ldots,z_{m,N}]^T \sim \mathcal {CN}(0,\sigma_z^2 \boldsymbol{I})$. After multiplying the time-domain signal by the DFT matrix, the received signal at the sensing receive antenna of the $m$-th BS in the frequency domain is given by
\begin{align}
 \bar{\boldsymbol{y}}_m=&[\bar{y}_{m,1},\ldots,\bar{y}_{m,N}]^T \nonumber \\ = & \boldsymbol{W} \boldsymbol{y}_m = \sqrt{p}\sum_{u=1}^M \mathrm{diag}(\boldsymbol{s}_u) \boldsymbol{G} \boldsymbol{h}_{u,m}+  \bar{\boldsymbol{z}}_m,\quad \forall m,
 \label{eqS2.9}
\end{align}
where $\boldsymbol{h}_{u,m}=[h_{u,m,1},\ldots,h_{u,m,L}]^T$; $\mathrm{diag}(\boldsymbol{s}_u)$ is a diagonal matrix with the main diagonal being $\boldsymbol{s}_u$; $\boldsymbol{G} \in \mathbb{C}^{N \times L} $ with the $(n,l)$-th element being $G_{n,l} = e^{\frac{-j2\pi (n-1)(l-1)}{N}}$; and $\bar{\boldsymbol{z}}_m =[\bar{z}_{m,1},\ldots,\bar{z}_{m,N}]^T\!=\! \boldsymbol{W} \boldsymbol{z}_m \!\sim\! \mathcal {CN}(0,\sigma_z^2 \boldsymbol{I})$ since $\boldsymbol{W} \boldsymbol{W}^H =\boldsymbol{I}$.

\section{Two-Phase Device-Free Sensing Framework}\label{sec:Two-Phase Target Sensing Framework}
At each coherence time, the received signal at the BS is given by (\ref{eqS2.9}) as mentioned above. To localize targets based on (\ref{eqS2.9}), we propose a two-phase framework for device-free sensing based on the signals received by the sensing receive antennas, as illustrated in Fig. \ref{fig12}.
\subsection{Phase I: Range Estimation}
First, in Phase I, each BS $m$ estimates the channels between its transmit antenna and its sensing receive antenna, $h_{m,m,l}$'s, $l=1,\ldots,L$. A key observation is that if ${h}_{m,m,l}\neq 0$ for some $l$, a target indexed by $\bar{k}_{m,l}\in \mathcal{K}$ exists, where the signal propagation from BS $m$ to target $\bar{k}_{m,l}$ and then back to BS $m$ experiences a delay of $l$ OFDM sample periods. Hence, by recalling that the duration of one OFDM sample period is $1/(N\Delta f)$ s, the distance (range) between target $\bar{k}_{m,l}$ and BS $m$ (which is half of the propagated distance) lies in the following range set (in m):
\begin{equation}
  \Theta(l)=\left \{d \left| \frac{(l-1) c_0}{2 N \Delta f}<d \leq \frac{l c_0}{2 N \Delta f} \right. \right\},
\label{eqS2.2}
\end{equation}
where $c_0$ denotes the speed of the light (in m/s). In this case, we propose to estimate the distance between the $m$-th BS and the $\bar{k}_{m,l}$-th target, $d_{m,\bar{k}_{m,l}}$, as the middle point in the above range set $\Theta(l)$:
\begin{equation}
 \bar{d}_{m,\bar{k}_{m,l}}=\frac{(l-1)c_0}{2N \Delta f} + \frac{c_0 }{4N \Delta f}, \text{ if } h_{m,m,l}\neq 0.
 \label{eqS2.10}
\end{equation}
Note that under the above estimation rule, the worst-case range estimation error is given by
\begin{equation}
 |\bar{d}_{m,\bar{k}_{m,l}}-{d}_{m,\bar{k}_{m,l}}|\leq \frac{c_0}{4 N \Delta f}\overset{\Delta}{=}\Delta d.
\label{eqS2.1}
\end{equation}
For example, in 5G OFDM systems, the channel bandwidth is $B=100$ MHz at the sub-6G frequency band and $B=400$ MHz at the mmWave band according to 3GPP Release 15 \cite{3gpp}. In this case, the worst-case range estimation errors are $0.75$ m and $0.1875$ m, respectively. Since $\Delta d$ is practically very small, we assume in the sequel that the values of distance for any two paths reflected back to a BS by two different targets differ by more than $2\Delta d$, thus the corresponding paths are resolvable. Therefore, there are $K$ non-zero entries in each $\boldsymbol{h}_{m,m}$.

After obtaining $\bar{d}_{m,\bar{k}_{m,l}}$'s, each BS $m$ has a range set consisting of the values of distance (ranges) with the targets:
\begin{equation}
\mathcal{D}_m=\{\bar{d}_{m,\bar{k}_{m,l}} |  \forall l \mbox{ satisfying } h_{m,m,l}>0\},\quad  \forall m.
\label{eqS2.11}
\end{equation}
For convenience, we define $\mathcal{D}_m(g)$ as the $g$-th largest element in $\mathcal{D}_m$, $\forall m$. Moreover, define $g_{m,k}$ as the mapping (or matching) between the element in $\mathcal{D}_m$ and the $k$-th target, such that $\bar{d}_{m,k}$ is the $g_{m,k}$-th largest element in $\mathcal{D}_m$, i.e.,
\begin{equation}
    \bar{d}_{m,k}=\mathcal{D}_m(g_{m,k}),\quad \forall m,k.
    \label{eqS2.12}
\end{equation}We call $g_{m,k}$'s as the \emph{data association} variables in the rest of this paper.

In Section \ref{sec:Range Estimation}, we will introduce in details the estimation of $h_{m,m,l}$'s based on the signals at the sensing receive antennas for obtaining the range sets $\mathcal{D}_{m}$'s in Phase I.
\subsection{Phase II: Localization Among BSs Based on Estimated Range}
Next, in Phase II, all the BSs share their range sets $\mathcal{D}_m$'s with each other via the cellular fronthaul links, and jointly estimate the location of each target $k$ based on the values of its distance to the $M$ BSs, $\{\bar{d}_{1,k},\ldots,\bar{d}_{M,k} \}$, $\forall k$. Note that in conventional device-based sensing for active targets that can send/receive RF signals with different signatures, the BSs can know the exact mapping between a range and a target, i.e., each BS $m$ knows which element in $\mathcal{D}_m$ belongs to target $k$, $\forall k$. However, in our considered device-free sensing for passive targets without communication capability, all the targets will reflect the same signals to the BSs. As a result, each BS $m$ only knows that the range of target $k$ lies in the set $\mathcal{D}_m$, but does not know which element in $\mathcal{D}_m$ corresponds to this range, i.e., $g_{m,k}$ is unknown, $\forall k$. In this case, a wrong data association solution of $g_{m,k}$'s may lead to the detection of \emph{ghost targets} (for which the exact definition will be given in Section \ref{sec:Localization with Perfect Range Estimation}) that do not exist, as illustrated in Example \ref{example1}.

\begin{example}
Suppose that there are $M=3$ BSs and $K=2$ targets. The coordinates of BSs 1, 2, and 3 are $(0,3)$, $(5,0)$, and $(0,-4)$, respectively, and the coordinates of targets 1 and 2 are $(2,-2)$ and $(-2,2)$, respectively. Suppose that the BSs can perfectly estimate the ranges of targets, i.e., $\bar{d}_{m,k}=d_{m,k}$, $\forall m, k$. Thus, we have $\mathcal{D}_1=\{\sqrt{29}, \sqrt{5}\}$, $\mathcal{D}_2=\{\sqrt{13},\sqrt{53}\}$, and $\mathcal{D}_3=\{2\sqrt{2}, 2\sqrt{10}\}$.

\begin{figure}
	\vspace{-5mm}
\begin{center}
  \scalebox{0.7}{\includegraphics*[29pt,16pt][385pt,307pt]{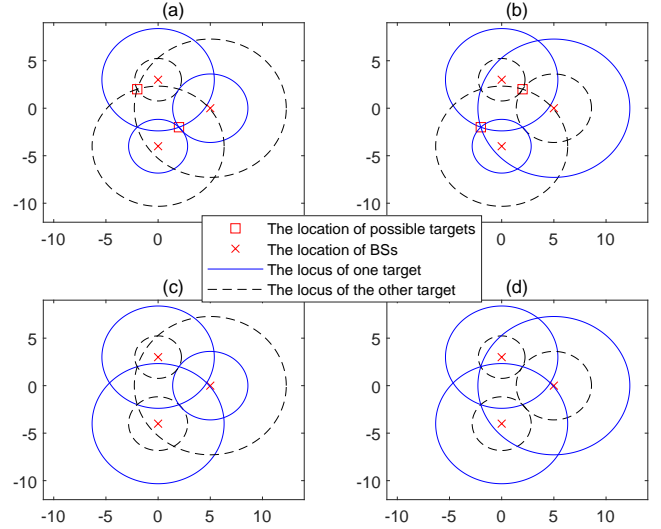}}
 \end{center}
 \caption{An example with ghost targets. In (a), we have $g_{1,1}=1$, $g_{2,1}=2$, $g_{3,1}=2$, $g_{1,2}=2$, $g_{2,2}=1$, $g_{3,2}=1$; in (b), we have $g_{1,1}=1$, $g_{2,1}=1$, $g_{3,1}=2$, $g_{1,2}=2$, $g_{2,2}=2$, $g_{3,2}=1$; in (c), we have $g_{1,1}=1$, $g_{2,1}=2$, $g_{3,1}=1$, $g_{1,2}=2$, $g_{2,2}=1$, $g_{3,2}=2$; in (d), we have $g_{1,1}=1$, $g_{2,1}=1$, $g_{3,1}=1$, $g_{1,2}=2$, $g_{2,2}=2$, $g_{3,2}=2$.}
\label{fig3}
\end{figure}

The job of the BSs is to solve the set of equations $\sqrt{(a_m-x_k)^2+(b_m-y_k)^2}=\mathcal{D}(g_{m,k})$, $m=1,2,3$, $k=1,2$, where $x_k$'s, $y_k$'s, and $g_{m,k}$'s are all unknown variables. If the BSs set $g_{1,1}=1$, $g_{2,1}=2$, $g_{3,1}=2$ for sensing one target, and $g_{1,2}=2$, $g_{2,2}=1$, $g_{3,2}=1$ for sensing the other target, i.e., $d_{1,1}=\mathcal{D}_1(1)=\sqrt{29}$, $d_{2,1}=\mathcal{D}_2(2)=\sqrt{13}$, and $d_{3,1}=\mathcal{D}_3(2)=2\sqrt{2}$ are used for sensing one target, and $d_{1,2}=\mathcal{D}_1(2)=\sqrt{5}$, $d_{2,2}=\mathcal{D}_2(1)=\sqrt{53}$, and $d_{3,2}=\mathcal{D}_3(1)=2\sqrt{10}$ are used for sensing the other target, then the real targets $(2,-2)$ and $(-2,2)$ can be detected, as shown in Fig. \ref{fig3} (a). However, if the BSs set $g_{1,1}=1$, $g_{2,1}=1$, $g_{3,1}=2$ for sensing one target and $g_{1,2}=2$, $g_{2,2}=2$, $g_{3,2}=1$ for sensing the other target, i.e., $d_{1,1}=\mathcal{D}_1(1)=\sqrt{29}$, $d_{2,1}=\mathcal{D}_2(1)=\sqrt{53}$, and $d_{3,1}=\mathcal{D}_3(2)=2\sqrt{2}$ are used for sensing one target, and $d_{1,2}=\mathcal{D}_1(2)=\sqrt{5}$, $d_{2,2}=\mathcal{D}_2(2)=\sqrt{13}$, and $d_{3,2}=\mathcal{D}_3(1)=2\sqrt{10}$ are used for sensing the other target, then two ghost targets with coordinates $(-2,-2)$ and $(2,2)$ will be detected, as shown in Fig. \ref{fig3} (b). Note that under the other data association solutions, e.g., $g_{1,1}=1$, $g_{2,1}=2$, $g_{3,1}=1$ for sensing one target and $g_{1,2}=2$, $g_{2,2}=1$, $g_{3,2}=2$ for sensing the other target, as shown in Fig. \ref{fig3} (c), and $g_{1,1}=1$, $g_{2,1}=1$, $g_{3,1}=1$ for one target and and $g_{1,2}=2$, $g_{2,2}=2$, $g_{3,2}=2$ for the other target, as shown in Fig. \ref{fig3} (d), no ghost target exists.
\label{example1}
\end{example}

Although Example \ref{example1} indicates that ghost targets may be detected under certain setups, it is worth noting that they do not always exist, as shown in the following example.

\begin{example}
Suppose that there are $M=3$ BSs and $K=2$ targets. The coordinates of BSs 1, 2, and 3 are $(0,3)$, $(5,0)$ and $(0,-4)$, respectively, and the coordinates of targets 1 and 2 are $(-1,2)$ and $(2,-1)$, respectively. Similar to Example \ref{example1}, suppose that the BSs can perfectly estimate the ranges of targets, i.e., $\mathcal{D}_1=\{ \sqrt{2}, 2\sqrt{5}\}$, $\mathcal{D}_2=\{2\sqrt{10},\sqrt{10}\}$, and $\mathcal{D}_3=\{\sqrt{37}, \sqrt{13}\}$. If the BSs set $g_{1,1}=1$, $g_{2,1}=2$, $g_{3,1}=2$ for sensing one target and $g_{1,2}=2$, $g_{2,2}=1$, $g_{3,2}=1$ for sensing the other target, as shown in Fig. \ref{fig4} (a), then the location of the real targets can be correctly estimated. Otherwise, with the other data association solutions of $g_{m,k}$'s, no ghost targets will be detected, as shown in Figs. \ref{fig4} (b), (c) and (d).

\begin{figure}
\vspace{-5mm}
\begin{center}
  \scalebox{0.7}{\includegraphics*[29pt,16pt][385pt,307pt]{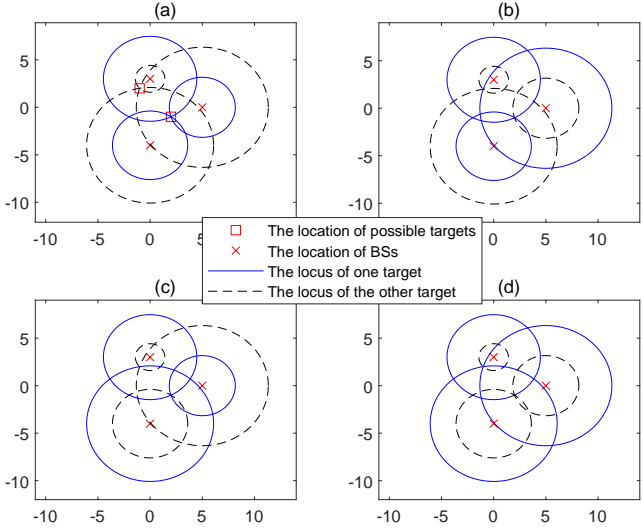}}
 \end{center}
 \caption{An example without ghost targets. In (a), we have $g_{1,1}=1$, $g_{2,1}=2$, $g_{3,1}=2$, $g_{1,2}=2$, $g_{2,2}=1$, $g_{3,2}=1$; in (b), we have $g_{1,1}=1$, $g_{2,1}=1$, $g_{3,1}=2$, $g_{1,2}=2$, $g_{2,2}=2$, $g_{3,2}=1$; in (c), we have $g_{1,1}=1$, $g_{2,1}=2$, $g_{3,1}=1$, $g_{1,2}=2$, $g_{2,2}=1$, $g_{3,2}=2$; in (d), we have $g_{1,1}=1$, $g_{2,1}=1$, $g_{3,1}=1$, $g_{1,2}=2$, $g_{2,2}=2$, $g_{3,2}=2$.}
\label{fig4}
\end{figure}

\label{example2}
\end{example}

In Section \ref{sec:Localization with Perfect Range Estimation}, we will study the fundamental limit of the ghost target detection probability arising from the data association issue at each BS; then, we will propose an ML algorithm to solve the joint data association and target localization problem to approach the above limit in Section \ref{sec:Localization with Imperfect Range Estimation}.

\section{Phase I: Range Estimation}\label{sec:Range Estimation}
In this section, we propose a range estimation algorithm for obtaining $\bar{d}_{m,\bar{k}_{m,l}}$'s as well as $\mathcal{D}_m$'s in Phase I, by estimating the multi-path channels $h_{m,m,l}$'s via OFDM channel estimation techniques.
\subsection{Algorithm Design}
Note that each BS $m$ only knows its own transmitted signals, i.e., $\boldsymbol{s}_m$, but does not know the transmitted signals of the other BSs, i.e., $\boldsymbol{s}_u$, $\forall u \neq m$. As a result, the main challenge to estimate the channels based on the frequency-domain received signal in (\ref{eqS2.9}) at each BS lies in the partial (instead of full) knowledge about the sensing matrix in (\ref{eqS2.9}). In the following, we show that the above challenge can be tackled by the fractional frequency reuse technique that is widely used in cellular networks to control the inter-cell interference \cite{5gstandard,Saquib13,Liu19}. Specifically, suppose that each BS $m$ occupies a partial of the $N$ sub-carriers denoted by the set $\mathcal{N}_m\subset \{1,\ldots,N\}$, i.e., \begin{align}
s_{m,n}=0, ~\forall n\notin \mathcal{N}_m, ~ \forall m.
\end{align}For simplicity, define $\mathcal{N}_m(n)$ as the $n$-th smallest element in the set $\mathcal{N}_m$. Moreover, adjacent BSs will be assigned with totally different sub-carrier sets under the fractional frequency reuse scheme. Let $\Upsilon_m$ denote the set of BSs that are far away from BS $m$ and thus share the same sub-carrier set with BS $m$, i.e., $\mathcal{N}_m=\mathcal{N}_u$ holds $\forall u\in \Upsilon_m$. Under the above scheme and according to (\ref{eqS2.9}), the received signal of BS $m$ at its assigned sub-carriers, denoted by $\tilde{\boldsymbol{y}}_m=[\bar{y}_{m,\mathcal{N}_m(1)},\ldots,\bar{y}_{m,\mathcal{N}_m(|\mathcal{N}_m|)}]^T$, is given by
\begin{equation}
 \tilde{\boldsymbol{y}}_m = \sqrt{p}\mathrm{diag}(\tilde{\boldsymbol{s}}_m) \tilde{\boldsymbol{G}}_m \boldsymbol{h}_{m,m}+  \tilde{\boldsymbol{z}}_m, ~ \forall m,
\label{eqS3.1}
\end{equation}where $\tilde{\boldsymbol{s}}_u=[s_{u,\mathcal{N}_u(1)},\ldots,s_{u,\mathcal{N}_u(|\mathcal{N}_u|)}]^T$, $\forall u\in \mathcal{M}$, $\tilde{\boldsymbol{G}}_m=[\boldsymbol{g}_{\mathcal{N}_m(1)},\ldots,\boldsymbol{g}_{\mathcal{N}_m(|\mathcal{N}_m|)}]^T$ with $\boldsymbol{g}_n^T$ being the $n$-th row of $\boldsymbol{G}$, and $\tilde{\boldsymbol{z}}_m=\sqrt{p}\sum_{u\in \Upsilon_m} \mathrm{diag}(\tilde{\boldsymbol{s}}_u) \tilde{\boldsymbol{G}}_m \boldsymbol{h}_{u,m}+  [\bar{z}_{m,\mathcal{N}_m(1)},\ldots,\bar{z}_{m,\mathcal{N}_m(|\mathcal{N}_m|)}]^T$ is the effective noise at BS $m$ with weak interference caused by the distant BSs in the set $\Upsilon_m$.

Note that among the $L$ elements in each $\boldsymbol{h}_{m,m}$, only $K\ll L$ of them are non-zero since there are merely $K$ scatters (targets) for each BS. Recovering the sparse channels $\boldsymbol{h}_{m,m}$'s based on (\ref{eqS3.1}) is thus a compressed sensing problem, which motivates us to use the least absolute shrinkage and selection operator (LASSO) \cite{LASSO} technique to estimate $\boldsymbol{h}_{m,m}$'s. Specifically, given a carefully designed parameter $\lambda\geq 0$,\footnote{More details on LASSO can be found in \cite{LASSO}.} the LASSO problem for estimating each $\boldsymbol{h}_{m,m}$ is formulated as
\begin{equation}
\begin{aligned}
 & \underset{{\boldsymbol{h}_{m,m}} }{\text{minimize}}
 ~ \frac{1}{2}\Vert \boldsymbol{\bar{y}}_{m,m} - \sqrt{p}\mathrm{diag}(\tilde{\boldsymbol{s}}_m) \tilde{\boldsymbol{G}}_m \boldsymbol{h}_{m,m} \Vert_2^2+ \lambda  \Vert \boldsymbol{h}_{m,m} \Vert_1.
  \end{aligned}
 \label{eqS3.2}
\end{equation}
Note that (\ref{eqS3.2}) is a convex optimization problem, for which the optimal solution can be efficiently obtained by CVX and serve as the estimated channel $\bar{\boldsymbol{h}}_{m,m}=[\bar{h}_{m,m,1},\ldots,\bar{h}_{m,m,L}]^T$.\footnote{For the special case of ${\rm rank}(\mathrm{diag}(\tilde{\boldsymbol{s}}_m) \tilde{\boldsymbol{G}}_m)=L$ (e.g., $N/M \geq L$) where  (\ref{eqS3.1}) describes an overdetermined linear system, we can set $\lambda=0$ in problem (\ref{eqS3.2}), which leads to the ML channel estimators $\bar{\boldsymbol{h}}_{m,m}=(\tilde{\boldsymbol{G}}_m^H\mathrm{diag}(\tilde{\boldsymbol{s}}_m)\mathrm{diag}(\tilde{\boldsymbol{s}}_m)\tilde{\boldsymbol{G}}_m)^{-1}\tilde{\boldsymbol{G}}_m^H\mathrm{diag}(\tilde{\boldsymbol{s}}_m)\tilde{\boldsymbol{y}}_m/\sqrt{p}$, $\forall m$.}

Then, if $\bar{h}_{m,m,l} \neq 0$ for some $l$, $\bar{d}_{m,\bar{k}_{m,l}}$ can be estimated via (\ref{eqS2.10}), and the range sets $\mathcal{D}_m$'s can be obtained via (\ref{eqS2.11}).

\subsection{Numerical Examples}
\begin{figure}
\centering
\includegraphics[scale=0.6]{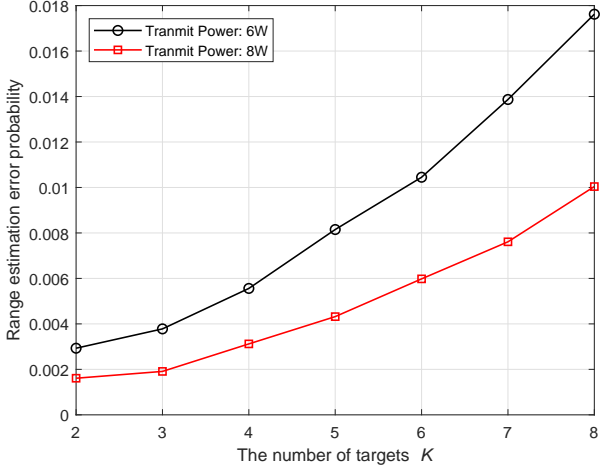}
\caption{Range estimation error probability versus the number of targets.}
\label{fig6}
\end{figure}

In the following, we provide a numerical example to evaluate the accuracy of the proposed range estimation algorithm. Specifically, we set $N=3300$ and $\Delta f=30$ kHz such that $B=100$ MHz \cite{3gpp}. According to \cite{zaidi2017nr}, with $\Delta f=30$ kHz, the length of the CP is $2.34$ $\mu$s. To make $L<Q$ such that the CP can be cancelled at the BSs, we assume that the maximum number of resolvable paths is $L=200$. Moreover, we consider $M=4$ BSs in the network, while each BS is randomly assigned with $N/M=825$ sub-carriers such that $\mathcal{N}_m \bigcap \mathcal{N}_u=\emptyset$, $\forall m \neq u$, i.e., $\Upsilon_m=\emptyset$, $\forall m$. Under this setup, we randomly generate $10^5$ independent localization realizations of the $M=4$ BSs and $K$ targets, following uniform distribution over a $200$ m $\times$ $200$ m square. Given the values of the BS-target distance, we can know the delay in terms of OFDM sample periods from BS $m$ to target $k$ back to BS $m$, $\forall m,k$, according to (\ref{eqS2.2}). Define $\mathcal{L}_m$ as the set of $K$ true delay values in terms of OFDM sample periods caused by the $K$ targets to BS $m$, $\forall m$. Then, we estimate the channels $\boldsymbol{h}_{m,m}$'s by solving problem (\ref{eqS3.2}), and define $\bar{\mathcal{L}}_m=\{l|\bar{h}_{m,m,l}\neq 0, l=1,\ldots,L\}$ as the set of estimated delay values at BS $m$, $\forall m$. If there exists at least an $m$ such that $\mathcal{L}_m\neq \bar{\mathcal{L}}_m$, we say that the range estimation is in error in this realization. Fig. \ref{fig6} shows the range estimation error probability versus the number of targets, $K$ (ranging from $2$ to $8$), where the BS transmit power is set as 6 Watt (W) and 8 W, respectively. It is observed that the range estimation error probability is very low under our proposed scheme, and can be significantly reduced by increasing the transmit power.

After $\bar{d}_{m,\bar{k}_{m,l}}$'s and the range sets $\mathcal{D}_m$'s are obtained, we need to study how the BSs can cooperate with each other to localize the $K$ targets based on $\mathcal{D}_m$'s. As discussed at the end of Section \ref{sec:Two-Phase Target Sensing Framework}, the main challenge here lies in the lack of information about $g_{m,k}$'s, i.e., each BS $m$ does not know how to match the ranges in $\mathcal{D}_m$ with the right targets, which may lead to the detection of ghost targets instead of the true targets. In the following, we will first consider the ideal case without error in estimating the values of the distance between the BSs and targets, i.e., $\bar{d}_{m,k}=d_{m,k}$, $\forall m, k$,  and investigate the fundamental limit of the ghost target detection probability in Section \ref{sec:Localization with Perfect Range Estimation}. Then, in Section \ref{sec:Localization with Imperfect Range Estimation}, we will focus on the practical case with possible error in estimating $d_{m,k}$'s, and propose an ML algorithm to find a good data association solution $g_{m,k}$'s at each BS so as to minimize the target localization error.

\section{Phase II: Localization with Perfect Range Estimation}\label{sec:Localization with Perfect Range Estimation}
In this section, we introduce the target localization in Phase II, assuming that the range estimation based on (\ref{eqS2.10}) is perfect, i.e., $\bar{d}_{m,k}=d_{m,k}$, $\forall m,k$, in order to derive the fundamental limits and draw essential insights. This corresponds to the ideal case of an infinite channel bandwidth such that $\Delta d=0$ in (\ref{eqS2.1}).

Note that in device-based sensing where each BS can distinguish the ranges of different targets, the locations of the targets can be estimated by solving the following equations:
\begin{equation}
	\sqrt{(a_m-x_k)^2+(b_m-y_k)^2}=d_{m,k},\quad \forall m, k.
	\label{eqS4.1}
\end{equation}
It is well-known that three BSs are sufficient to localize the targets. However, in our considered device-free sensing where each BS cannot distinguish the ranges of different targets, the BSs need to solve the following equations to estimate the target locations:
\begin{align}
	& \sqrt{(a_m-x_k)^2+(b_m-y_k)^2}=\mathcal{D}_m(g_{m,k}),\quad \forall m, k,
	\label{eqS4.2} \\
	& g_{1,k}=k, \quad \forall k, \label{eqn:BS 1} \\
	& \{g_{m,1},\ldots,g_{m,K}\}=\mathcal{K}, \quad \forall m>2. \label{eqn:matching}
\end{align}
Different from the device-based sensing equations in (\ref{eqS4.1}), under device-free sensing, $g_{m,k}$'s are also unknown variables in (\ref{eqS4.2}) that need to satisfy conditions (\ref{eqn:BS 1}) and (\ref{eqn:matching}), i.e., different ranges for each BS belong to different targets. In (\ref{eqn:BS 1}), we define target $k$ as the target whose distance to BS $1$ is the $k$-th largest element in $\mathcal{D}_1$, i.e., $d_{1,k}=\mathcal{D}_1(k)$, $\forall k\in \mathcal{K}$. The reason is to mitigate the ambiguity in target indexing. To illustrate this, let us consider Example \ref{example1} in Section \ref{sec:Range Estimation}. In Fig. \ref{fig3} (a), under the data association solution of $g_{1,1}=1$, $g_{2,1}=2$, $g_{3,1}=2$, $g_{1,2}=2$, $g_{2,2}=1$, $g_{3,2}=1$, the location of target 1 is $(2,-2)$, while that of target 2 is $(-2,2)$. However, if we consider the data association solution of $g_{1,1}=2$, $g_{2,1}=1$, $g_{3,1}=1$, $g_{1,2}=1$, $g_{2,2}=2$, $g_{3,2}=2$, the location of target 1 is $(-2,2)$, while that of target 2 is $(2,-2)$. In fact, these two data association solutions lead to the same localization result. Thus, we add constraint (\ref{eqn:BS 1}) to avoid the above ambiguity.

It is worth noting that in device-free sensing, the target locations may not be accurately estimated by three BSs because there may be multiple solutions to equations (\ref{eqS4.2}), (\ref{eqn:BS 1}), and (\ref{eqn:matching}) when $M=3$, which leads to the existence of \emph{ghost targets}, as shown in Example 1. In the following, we propose an algorithm to detect the existence of ghost targets, and derive the fundamental limit of ghost target existence.
\subsection{Definition of Ghost Targets}
First, we present the rigourous definition of ghost targets.

\begin{definition}
Define $\mathcal{X}=\{(x_1,y_1),\ldots,(x_K,y_K)\}$ as the set of coordinates for all the $K$ targets. Then, consider another coordinate set $\mathcal{X}^{\rm G}=\{(x_1^{\rm G},y_1^{\rm G}),\ldots,(x_K^{\rm G},y_K^{\rm G})\}\neq \mathcal{X}$ and define
\begin{align}
&d_{m,k}^{\rm G}=\sqrt{(a_m-x_k^{\rm G})^2+(b_m-y_k^{\rm G})^2},\quad \forall m, k, \label{eqS4.4}
 \\
&\mathcal{D}_m^{\rm G}=\{ d_{m,1}^{\rm G},\ldots,d_{m,K}^{\rm G} \},\quad \forall m. \label{eqS4.5}
\end{align}
We say $K$ ghost targets exist with the coordinates shown in the set $X^{\rm G}$ if and only if
\begin{equation}
\mathcal{D}_m=\mathcal{D}_m^{\rm G}, \quad \forall m.
\label{eqS4.6}
\end{equation}
In other words, besides $\mathcal{X}$, $\mathcal{X}^{{\rm G}}$ is another solution to equations (\ref{eqS4.2}), (\ref{eqn:BS 1}), and  (\ref{eqn:matching}).
\label{definition1}
\end{definition}

\subsection{Algorithm to Detect the Existence of Ghost Targets}
Based on Definition \ref{definition1}, given any particular BS locations $(a_m,b_m)$'s and range sets $\mathcal{D}_m$'s, we can efficiently check whether ghost targets exist as follows. First, since three BSs can locate any target if the data association solution is given, we can fix the data association solution for BS 1 as (\ref{eqn:BS 1}) and list all the feasible data association solutions for BSs 2 and 3 that satisfy condition (\ref{eqn:matching}). In total, there are $(K!)^2$ feasible data association solutions for BSs 2 and 3. Moreover, a feasible data association solution for BSs 1, 2, and 3 should also satisfy the following triangle inequalities for each target $k$:
\begin{align}
  |\mathcal{D}_{m_1}(g_{m_1,k})-\mathcal{D}_{m_2}(g_{m_2,k})|\leq d_{m_1,m_2}^{\rm{BS}}, ~ \forall m_1,m_2\in \{1,2,3\}, \label{eqS5.60}\\
  |\mathcal{D}_{m_1}(g_{m_1,k})+\mathcal{D}_{m_2}(g_{m_2,k})|\geq d_{m_1,m_2}^{\rm{BS}}, ~ \forall m_1,m_2\in \{1,2,3\}.\label{eqS5.70}
\end{align}To summarize, we can define a set consisting of all the feasible data association solutions for BSs 1, 2, and 3 as follows:
\begin{align}\label{eqS5.80}
\mathcal{H}=&\{\{g_{1,k},g_{2,k},g_{3,k}\}_{k=1}^K|\{g_{1,k},g_{2,k},g_{3,k}\}_{k=1}^K ~ {\rm satisfies} ~ (\ref{eqn:BS 1}), \nonumber \\ & (\ref{eqn:matching}), ~ {\rm and} ~ (g_{1,k},g_{2,k}), ~ (g_{1,k},g_{3,k}), ~ (g_{2,k},g_{3,k}) ~ {\rm satisfy} \nonumber \\ & (\ref{eqS5.60}), ~ (\ref{eqS5.70}), ~ \forall k\}.
\end{align}
Usually, the cardinality of $\mathcal{H}$ is much smaller than $(K!)^2$ thanks to the utilization of the triangle inequalities (\ref{eqS5.60}) and (\ref{eqS5.70}) to eliminate the infeasible data association solutions.

Then, for each data association solution for BSs 1, 2, and 3 in the set $\mathcal{H}$, which is given by $\bar{g}_{m,k}$'s, $m=1, 2, 3$ and $\forall k$, we check whether there is a localization solution to the following equations:
\begin{equation}
\sqrt{(a_m-x_k)^2+(b_m-y_k)^2}=\mathcal{D}_m(\bar{g}_{m,k}), m=1, 2, 3, ~ \forall k.
\label{eqS4.7}
\end{equation}
If there is no solution, we can conclude that under all the data association solutions for all the BSs where the data association solution to BSs 1, 2, and 3 is $g_{m,k}=\bar{g}_{m,k}$, $m=1,2,3$ and $k=1,\ldots,K$, no ghost target exists. Otherwise, if there is a solution denoted by $\{(\bar{x}_k,\bar{y}_k)\}_{k=1}^K$ to the above equations, we can use this solution to calculate $\bar{\mathcal{D}}_m$, $\forall m>3$, the elements of which are $\sqrt{(a_m-\bar{x}_k)^2+(b_m-\bar{y}_k)^2}$'s, $k\in \mathcal{K}$. If there exists an $\bar{m}>3$ such that $\bar{\mathcal{D}}_{\bar{m}} \neq {\mathcal{D}}_{\bar{m}}$, then we can conclude that under all the data association solutions for all the BSs where the data association solution to BSs 1, 2, and 3 is $g_{m,k}=\bar{g}_{m,k}$, $m=1,2,3$ and $k=1,\ldots,K$, no ghost targets exist. Otherwise, if $\bar{\mathcal{D}}_m=\mathcal{D}_m$, $\forall m>3$, then $\{(\bar{x}_k,\bar{y}_k)\}_{k=1}^K$ is defined as a feasible localization solution, which may be the locations of either the true targets or the ghost targets. After searching over all the feasible data association solutions for BSs 1, 2, and 3 in the set $\mathcal{H}$, if only one feasible solution of  $\{(\bar{x}_k,\bar{y}_k)\}_{k=1}^K$ is found, it indicates that given this particular $(a_m,b_m)$'s and $\mathcal{D}_m$'s, no ghost target exists. Otherwise, if multiple feasible solutions of $\{(\bar{x}_k,\bar{y}_k)\}_{k=1}^K$ are found, then ghost targets exist given this $(a_m,b_m)$'s and $\mathcal{D}_m$'s. A summary of this algorithm is given in Algorithm \ref{alg:1}.

\begin{algorithm}[t]
{\bf Input}: $(a_m,b_m)$'s and $\mathcal{D}_m$'s, $\forall m \in \mathcal{M}$. \\
{\bf Initialization}: Obtain the feasible data association solutions for BSs 1, 2, and 3 in $\mathcal{H}$ given in (\ref{eqS5.80}). Define $\mathcal{H}(t)$ as the $t$-th data association solution in $\mathcal{H}$. Set $t=1$ and $\tau=0$. \\
{\bf Repeat}:
\begin{enumerate}
\item[1.] Set $\mathcal{H}(t)$ as the data association solution of BSs 1, 2, and 3, denoted by $\bar{g}_{m,k}$'s, $m=1, 2, 3$ and $k=1, \ldots, K$;
\item[2.] Check whether there exists a localization solution to equations (\ref{eqS4.7}) given the above data association solution. If there exists a solution, which is denoted by $\{(\bar{x}_{k},\bar{y}_{k})\}_{k=1}^K$, then:
\begin{enumerate}
\item[2.1] For each BS $m>3$, set $\bar{\mathcal{D}}_m=\{\sqrt{(a_m-\bar{x}_{k})^2+(b_m-\bar{y}_{k}^2)}|k=1,\ldots,K\}$;
\item[2.2] Check whether $\bar{\mathcal{D}}_m=\mathcal{D}_m$, $\forall m>3$, holds. If this is true, set $\tau=\tau+1$.
\end{enumerate}
\item[3.] Set $t=t+1$.
\end{enumerate}
{\bf Until} $t=|\mathcal{H}|$. \\
{\bf Output}: If $\tau=1$, no ghost target exists; otherwise, if $\tau>1$, ghost target exists.
\caption{Algorithm to Check Whether Ghost Target Exists}
\label{alg:1}
\end{algorithm}

\subsection{Fundamental Limit for Existence of Ghost Target}
Note that Algorithm \ref{alg:1} can help us determine whether ghost target exists given any $(a_m,b_m)$'s and $\mathcal{D}_m$'s. In the following, we aim to show some stronger results about the fundamental limit of the ghost target detection probability merely given the BS locations, i.e., $(a_m,b_m)$'s, but regardless of the range sets $\mathcal{D}_m$'s. To achieve this goal, given any target coordinate set $\mathcal{X}$ and another set $\mathcal{X}^{{\rm G}}\neq \mathcal{X}$ consisting of $K$ pairs of coordinates, define
\begin{align}
& \mathcal{X}^{{\rm C}}= \mathcal{X} \bigcap  \mathcal{X}^{{\rm G}}, \label{eqn:common set} \\
& \tilde{\mathcal{X}}=\mathcal{X}/\mathcal{X}^{{\rm C}}, \label{eqn:set X} \\
& \tilde{\mathcal{X}}^{{\rm G}}=\mathcal{X}^{{\rm G}}/ \mathcal{X}^{{\rm C}},
\label{eqn:set X1}
\end{align}where $\mathcal{A}/\mathcal{B}=\{x|x\in \mathcal{A} ~ {\rm and} ~ x\notin \mathcal{B}\}$. As a result, $\mathcal{X}^{{\rm C}}$ is the set of common coordinates in $\mathcal{X}$ and $\mathcal{X}^{{\rm G}}$, while $\tilde{\mathcal{X}}$ and $\tilde{\mathcal{X}}^{{\rm G}}$ consist of the distinct parts in $\mathcal{X}$ and $\mathcal{X}^{{\rm G}}$. Then, define
\begin{equation}
\mathcal{S}_{k,q}=\{m|d_{m,k}=d_{m,q}^{{\rm G}}, \forall m\},\quad \forall k,q,
\label{eqS4.8}
\end{equation}
as the set of BSs whose distance values to $(x_k,y_k)$ and $(x_q^{\rm G},y_q^{\rm G})$ are the same, where $d_{m,q}^{{\rm G}}$ is given in (\ref{eqS4.4}). Note that if $(x_k,y_k)=(x_q^{{\rm G}},y_q^{{\rm G}})\in \mathcal{X}^{{\rm C}}$, then $\mathcal{S}_{k,q}=\mathcal{M}$. Otherwise, if $(x_k,y_k)\in \tilde{\mathcal{X}}$ and $(x_q^{{\rm G}},y_q^{{\rm G}})\in \tilde{\mathcal{X}}^{{\rm G}}$ such that $(x_k,y_k)\neq (x_q^{{\rm G}},y_q^{{\rm G}})$, then all the BSs in the set $\mathcal{S}_{k,q}$ must be on the perpendicular bisector of the line segment connecting $(x_k,y_k)$ and $(x_q^{\rm G},y_q^{\rm G})$. In the following, we provide one BS deployment topology where ghost target never exists no matter where the true targets are.

\begin{theorem}\label{Theorem1}
Suppose that range estimation is perfect at all BSs. If $M \geq 2K+1$ and any three of the BSs are not deployed on the same line, then no matter where the $K$ targets are, ghost target does not exist.
\end{theorem}

\begin{IEEEproof}
We prove Theorem \ref{Theorem1} by contradiction. Suppose that there exists a coordinate set for the $K$ true targets $\mathcal{X}=\{(x_1,y_1),\ldots,(x_K,y_K)\}$ such that $K$ ghost targets exist with a coordinate set $\mathcal{X}^{{\rm G}}=\{(x_1^{{\rm G}},y_1^{{\rm G}}),\ldots,(x_K^{{\rm G}},y_K^{{\rm G}})\}\neq \mathcal{X}$. Let us consider a coordinate $(x_k,y_k)\in \mathcal{X}$ which however does not appear in $\mathcal{X}^{{\rm G}}$, i.e., $(x_k,y_k)\in \tilde{\mathcal{X}}$ defined in (\ref{eqn:set X}). In this case, the BSs in the set $\mathcal{S}_{k,q}$ should be on the perpendicular bisector of the line segment connecting $(x_k,y_k)$ and $(x_q^{{\rm G}},y_q^{{\rm G}})$, $\forall q\in \mathcal{K}$. Since any three BSs are not deployed on the same line, we have $|\mathcal{S}_{k,p}|=0$, $1$, or $2$, $\forall q\in \mathcal{K}$. It thus follows that $\sum\limits_{q=1}^K|\mathcal{S}_{k,q}|\leq 2K<M$. In other words, $\bigcup_{q \in \mathcal{K}} \mathcal{S}_{k,q} \neq \mathcal{M}$, and there exists an $m\in \mathcal{M}$ such that $m\notin \bigcup_{q\in \mathcal{K}}\mathcal{S}_{k,q}$. This indicates that $d_{m,k}$ is not in the set $\mathcal{D}_{m}^{{\rm G}}$ and $\mathcal{D}_{m}\neq \mathcal{D}_{m}^{{\rm G}}$, which contradicts (\ref{eqS4.6}) in Definition \ref{definition1}. Therefore, if $M\geq 2K+1$ and any three BSs are not deployed on the same line, then there never exist ghost targets no matter where the $K$ true targets are. Theorem \ref{Theorem1} is thus proved.
\end{IEEEproof}

The key condition for Theorem \ref{Theorem1} is $M\geq 2K+1$. It is worth noting that if this condition does not hold, there may exist some target locations that can lead to ghost targets, as shown in Example \ref{example1}, where $M=3$ and $K=2$. Interestingly, the following theorem shows that even if $M<2K+1$, when the targets are located independently and uniformly in the network, the probability that these targets happen to be at the locations that can lead to ghost targets is \emph{zero}.

\begin{theorem}\label{Theorem2}
Suppose that range estimation is perfect at all BSs. If $M\geq 4$ and any three of the BSs are not deployed on the same line, then given any finite number of targets, ghost target does not exist almost surely when the true targets are located independently and uniformly in the network.
\end{theorem}

\begin{IEEEproof}
Please refer to Appendix \ref{appendix1}.
\end{IEEEproof}

In the following, we provide a toy example with $M=4$ BSs and $K=2$ targets to help understand Theorem \ref{Theorem2}.

\begin{lemma}
Suppose that range estimation is perfect at all BSs. Consider the case of $M =4$ and $K=2$, where any three of the BSs are not deployed on the same line. If the line connecting any two BSs is not perpendicular to the line connecting the other two BSs, then no matter where the $K$ true targets are, there never exist ghost targets. If there exist two BSs such that the line connecting them is perpendicular to the line connecting the other two BSs with an intersection point $(x_0,y_0)$, then there exist ghost targets only when the coordinates of the two true targets satisfy $x_1+x_2=2x_0$ and $y_1+y_2=2y_0$, i.e., the intersection point $(x_0,y_0)$ is the middle point of the line segment connecting the two true targets.
\label{Lemma1}
\end{lemma}

\begin{IEEEproof}
Please refer to Appendix \ref{appendix2}.
\end{IEEEproof}

\begin{figure}
\begin{center}
\subfigure[Never without ghost targets]
{\scalebox{0.6}{\includegraphics*{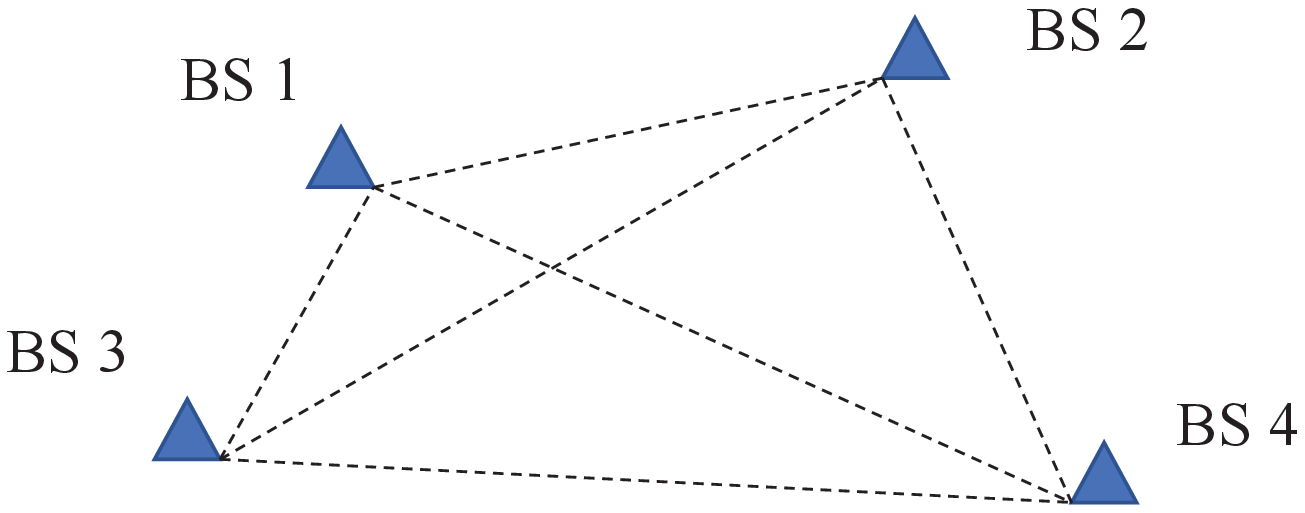}}}
\subfigure[Without ghost targets almost surely]
{\scalebox{0.75}{\includegraphics*{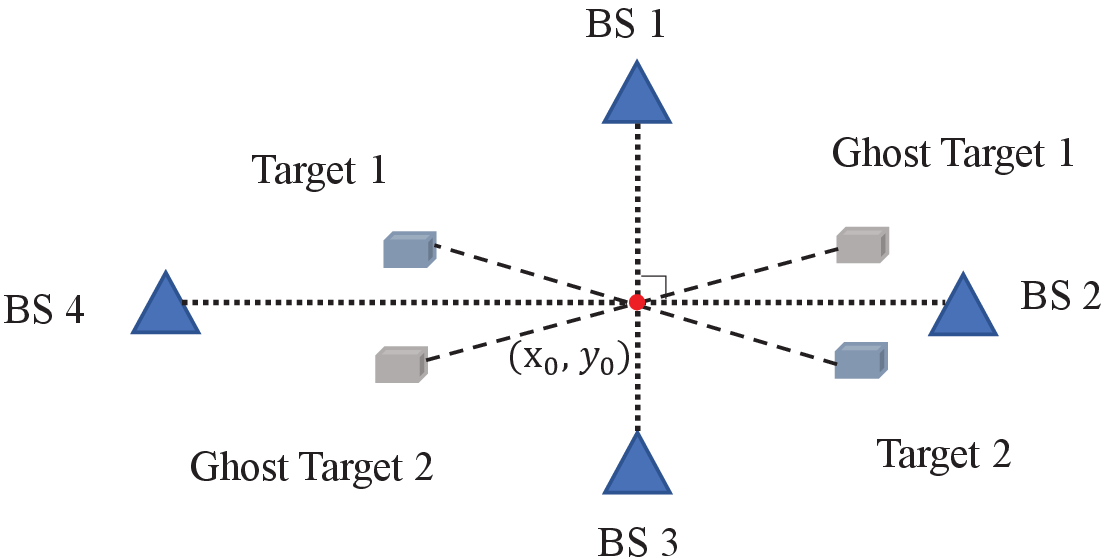}}}
\end{center}
\caption{Examples of the BS deployment strategies shown in Lemma \ref{Lemma1} for detecting $K=2$ targets with $M=4$ BSs.}\label{fig15}
\end{figure}

The BS deployment strategies of Lemma \ref{Lemma1} are shown in Fig. \ref{fig15} with $M=4$ BSs, where ghost targets never exist regardless of the location of the $K=2$ true targets under the strategy shown in Fig. \ref{fig15}(a), and may exist given some special target location, i.e., $x_1+x_2=2x_0$ and $y_1+y_2=2y_0$, under the strategy shown in Fig. \ref{fig15}(b). Lemma \ref{Lemma1} indicates that even if the BSs are deployed as in Fig. \ref{fig15} (b), ghost targets do not exist almost surely  if the targets are located independently and uniformly in the network, because the probability of the event of $x_1+x_2=2x_0$ and $y_1+y_2=2y_0$ is zero in a four-dimension space constructed by $x_1,x_2,y_1,y_2$.

We also implement tremendous Monte Carlo simulations to verify Theorem \ref{Theorem2}, where we set $M=4$ and $K$ up to $20$. For each value of $K$, we generate $10^5$ realizations, where BS and target locations are generated independently and randomly under the uniform distribution in each realization, similar to the setup for Fig. \ref{fig6}. Based on the generated locations of BSs and targets, we use Algorithm \ref{alg:1} to check whether ghost target exists. It is observed that no ghost target is detected in any realization.

Theorems \ref{Theorem1} and \ref{Theorem2} provide a theoretical guarantee to the performance of our proposed two-phase device-free sensing framework. Specifically, compared to device-based sensing, a key issue in device-free sensing is the potential existence of ghost targets as pointed out in the above. However, Theorems \ref{Theorem1} and \ref{Theorem2} imply that this issue is actually not the bottleneck under our considered framework, since the ghost target never exists, or does not exist almost surely, depending on the relationship between the number of BSs and that of the targets.

Despite the above results, data association is a main technical challenge for implementing device-free sensing compared to implementing device-based sensing, although such an association will not degrade the fundamental performance as explained above. In the next section, we study the joint data association and localization algorithm in Phase II under the practical case with imperfect range estimation in Phase I.

\section{Phase II: Localization with Imperfect Range Estimation}\label{sec:Localization with Imperfect Range Estimation}
In this section, we consider the practical case when the range estimation is imperfect in Phase I due to the limited bandwidth, i.e., $\bar{d}_{m,k}$ shown in (\ref{eqS2.10}) is not equal to $d_{m,k}$, $\forall m, k$. In this case, we propose an ML-based algorithm to estimate the locations of the targets based on the knowledge of the BS locations, i.e., $(a_m,b_m)$'s, and the target range sets, i.e., $\mathcal{D}_m$'s.
\subsection{Algorithm Design}
 Similar to \cite{aditya2017localization,Anderson07}, in the rest of this paper, we assume that the estimated range shown in (\ref{eqS2.10}) follows
\begin{equation}
\bar{d}_{m,k}=d_{m,k}+\epsilon_{m,k}, \forall m,k,
\label{eqS5.1}
\end{equation}
where $\epsilon_{m,k} \sim \mathcal{CN}(0,\sigma_{m,k}^2)$ denotes the error for estimating $d_{m,k}$ and $\epsilon_{m,k}$'s are independent over $m$ and $k$. Based on the above range estimation model, for each BS $m$ and target $k$, the conditional probability for the event that $\bar{d}_{m,k}$ is the $g_{m,k}$-th largest element in $\mathcal{D}_m$, i.e., $\bar{d}_{m,k}=\mathcal{D}_m(g_{m,k})$, given $(x_k,y_k)$ and $g_{m,k}$, is
\begin{align}
&f_{m,k}(\mathcal{D}_m(g_{m,k})|x_k,y_k,g_{m,k}) \nonumber \\
=&\frac{1}{\sqrt{2\pi}\sigma_{m,k}} e^{-\frac{(\mathcal{D}_m(g_{m,k})-\sqrt{(a_m-x_k)^2+(b_m-y_k)^2})^2}{2\sigma_{m,k}^2}},~ \forall m,k.
\label{eqS5.2}
\end{align}

Define $\mathcal{G}=\{g_{m,k}|m=1,\ldots,M,k=1,\ldots,K \}$. Then, the ML problem to estimate the locations of all the $K$ targets can be formulated as
\begin{align}
\mathop{\textup{maximize}}_{\mathcal{X},\mathcal{G}}&~\prod_{k=1}^K \prod_{m=1}^M f_{m,k}(\mathcal{D}_m(g_{m,k})|x_k,y_k,g_{m,k})
\label{eqS5.3} \\
\textup{subject to}&~ (\ref{eqn:BS 1}), (\ref{eqn:matching}). \nonumber
\end{align}
Note that different from the device-based sensing problem, the data association solutions $g_{m,k}$'s need to be jointly optimized with the target locations in our considered device-free sensing problem, since each BS $m$ does not know the matching between the elements in $\mathcal{D}_m$ and the targets.

Based on (\ref{eqS5.2}), problem (\ref{eqS5.3}) can be simplified into the following problem
\begin{align}
\mathop{\textup{minimize}}_{\mathcal{X},\mathcal{G}}&
\sum_{k=1}^K\sum_{m=1}^M \frac{(\mathcal{D}_m(g_{m,k})\!-\!\sqrt{(a_m\!-\!x_k)^2+(b_m\!-\!y_k)^2})^2}{\sigma_{m,k}^2}
 \label{eqS5.4} \\
\textup{subject to}& ~ (\ref{eqn:BS 1}), (\ref{eqn:matching}). \nonumber
\end{align}Problem (\ref{eqS5.4}) is a non-convex optimization problem, which is thus difficult to solve. Nevertheless, it is worth noting that given any data association solution satisfying conditions (\ref{eqn:BS 1}) and (\ref{eqn:matching}), denoted by $g_{m,k}=\bar{g}_{m,k}$, $\forall m,k$, problem (\ref{eqS5.4}) can be decoupled into $K$ subproblems, each being formulated as follows for estimating the location of target $k$:
\begin{align}
\mathop{\textup{minimize}}_{x_k,y_k}&~
\sum_{m=1}^M \frac{(\mathcal{D}_m(\bar{g}_{m,k})-\sqrt{(a_m-x_k)^2+(b_m-y_k)^2})^2}{\sigma_{m,k}^2}.
\label{eqS5.5}
\end{align}
Similar to the device-based localization scenario, problem (\ref{eqS5.5}) given the data association solution is a nonlinear least squared problem, which can be solved efficiently by using the Gauss-Newton algorithm \cite{torrieri1984statistical,gavish1992performance}. As a result, problem (\ref{eqS5.4}) can be solved in a straightforward manner based on the exhaustive search method. Specifically, given any data association solution $g_{m,k}=\bar{g}_{m,k}$'s satisfying (\ref{eqn:BS 1}) and (\ref{eqn:matching}), we can solve problem (\ref{eqS5.5}) to obtain the locations of all the targets and check the corresponding objective value of problem (\ref{eqS5.4}). Then, after all the data association solutions satisfying (\ref{eqn:BS 1}) and (\ref{eqn:matching}) are searched, we can select the data association solution that minimizes the objective function of problem (\ref{eqS5.4}). However, the above approach based on exhaustive search is of prohibitively high complexity in practice. Specifically, there are $(K!)^{M-1}$ different data association solutions for $g_{m,k}$'s satisfying (\ref{eqn:BS 1}) and (\ref{eqn:matching}). Moreover, given each feasible data association solution, we need to solve the complex optimization problem (\ref{eqS5.5}) for $K$ times (each corresponding to one target). This thus motivates us to propose a low-complexity algorithm for solving problem (\ref{eqS5.4}).

To this end, we first note that some data association solutions can be easily determined to be infeasible with very high probability under imperfect range estimation. For instance, for any two range sets $\mathcal{D}_{m_1}$ and $\mathcal{D}_{m_2}$, if  $\mathcal{D}_{m_1}(g_{m_1,k})$ and $\mathcal{D}_{m_2}(g_{m_2,k})$ do not satisfy the triangle inequalities for target $k$, i.e.,
\begin{align}
  |\mathcal{D}_{m_1}(g_{m_1,k})-\mathcal{D}_{m_2}(g_{m_2,k})|\leq d_{m_1,m_2}^{\rm{BS}}+\delta_0 , ~ \forall m_1,m_2, \label{eqS5.6}\\
  |\mathcal{D}_{m_1}(g_{m_1,k})+\mathcal{D}_{m_2}(g_{m_2,k})|\geq d_{m_1,m_2}^{\rm{BS}}-\delta_0 , ~ \forall m_1,m_2,\label{eqS5.7}
\end{align}
where $\delta_0 > 0$ is some given value, then $(g_{m_1,k},g_{m_2,k})$ is not a feasible data association solution for target $k$ with very high probability. Note that different from (\ref{eqS5.60}) and (\ref{eqS5.70}) for the case of perfect range estimation, we put a margin $\delta_0$ here considering the imperfect estimation of $d_{m,k}$'s.

Inspired by the above property, we propose a low-complexity algorithm to solve problem (\ref{eqS5.4}) as follows. First, we just consider BSs 1, 2, and 3. Define the set of feasible data association solutions for these 3 BSs as
\begin{align}
\mathcal{G}^{(3)}=&\{\{g_{1,k},g_{2,k},g_{3,k}\}_{k=1}^K|\{g_{1,k},g_{2,k},g_{3,k}\}_{k=1}^K ~ \text{satisfies}  \nonumber \\ & (\ref{eqn:BS 1}), ~ (\ref{eqn:matching}), ~ \text{and} ~ (g_{1,k},g_{2,k}),~ (g_{1,k},g_{3,k}), ~ (g_{2,k},g_{3,k}) \nonumber \\ &  \text{satisfy} ~ (\ref{eqS5.6}), ~ (\ref{eqS5.7}), ~ \forall k\}.
\label{eqS5.8}
\end{align}

Next, given any $\{\bar{g}_{1,k},\bar{g}_{2,k},\bar{g}_{3,k}\}_{k=1}^K \in \mathcal{G}^{(3)}$, we solve problem (\ref{eqS5.5}) by setting $M=3$ to find the location of target $k$, $\forall k$. Let $(x_k^{(3)},y_k^{(3)})$'s, $k=1,\ldots,K$, denote the obtained solutions. Then, given these solutions, we can check the distance from any target $k$ to any BS $m>3$ as
\begin{align}
\bar{d}_{m,k}^{(3)}=\sqrt{(a_m-x_k^{(3)})^2+(b_m-y_k^{(3)})^2}.
\end{align}
Note that for any BS $m>3$, it does not know which element in $\mathcal{D}_m$ is the distance of target $k$ to it. If BS $m>3$ decides that $g_{m,k}=\tilde{k}$, then we define a cost for this decision as
\begin{equation}
\Delta d_{m,k,\tilde{k}}=|\mathcal{D}_m(\tilde{k})-\bar{d}_{m,k}^{(3)}|,~
\forall k,\tilde{k}, ~ \text{and} ~ m>3.
 \label{eqS5.11}
\end{equation}
As a result, the cost for BS $m>3$ to select $g_{m,k}=\tilde{k}$ is the error for using $\bar{d}_{m,k}^{(3)}$ to replace $\mathcal{D}_m(\tilde{k})$.

Define the indicator functions for matching as follows:
\begin{equation}
 \beta_{m,k,\tilde{k}}=
 \left \{
 \begin{array}{ll}
1,&\text{if $g_{m,k}$ is set to be $\tilde{k}$,}  \\
0,&\text{otherwise},
\end{array}
 \right.
 \forall k, \tilde{k}, m>3.
\label{eqS5.12}
\end{equation}
Note that for each BS $m>3$, (\ref{eqn:matching}) indicates that any $\tilde{k}$ can only be assigned to one target. Moreover, for any target $k$, only one $\tilde{k}$ can be assigned to it. As a result, we have the following constraints for the indicator functions:
\begin{align}
&\sum_{k=1}^K \beta_{m,k,\tilde{k}}=1, \quad \forall \tilde{k} ~ \text{and} ~ m>3, \label{eqS5.13}\\
&\sum_{\tilde{k}=1}^K \beta_{m,k,\tilde{k}}=1, \quad \forall k ~ \text{and} ~ m>3. \label{eqS5.14}
\end{align}

Define $\mathcal{B}_m=\{\beta_{m,k,\tilde{k}}|\forall k, \tilde{k} \in \mathcal{K} \}$, $\forall m>3$. Given the estimated target locations $(x_k^{(3)},y_k^{(3)})$'s, we aim to find a data association solution for each BS $m>3$ such that the overall mismatch between $\bar{d}_{m,k}$'s and $\bar{d}_{m,k}^{(3)}$'s for all the targets is minimized, for which we formulate the following optimization problem for any BS $m>3$:
\begin{align}
\mathop{\textup{minimize}}_{\mathcal{B}_m}&~ \sum_{k=1}^K \sum_{\tilde{k}=1}^K \beta_{m,k,\tilde{k}} \Delta d_{m,k,\tilde{k}} \label{eqS5.15} \\
\textup{subject to}&~ (\ref{eqS5.13}), (\ref{eqS5.14}). \nonumber
\end{align}
Problem (\ref{eqS5.15}) is an \emph{assignment problem}, which can be efficiently solved by the Hungarian algorithm \cite{kuhn1955hungarian}. After solving problem (\ref{eqS5.15}) for all the BSs $m>3$, we can find the data association solutions of $g_{m,k}$'s for these BSs based on (\ref{eqS5.12}).

Given any feasible data association solutions for BSs 1, 2, and 3 denoted by $\{\bar{g}_{1,k},\bar{g}_{2,k},\bar{g}_{3,k}\}_{k=1}^K\in \mathcal{G}^{(3)}$, after the assignment problem (\ref{eqS5.15}) is solved for the other BSs, the data association solutions for all the $M$ BSs, denoted by $\{\bar{g}_{1,k},\ldots,\bar{g}_{M,k}\}_{k=1}^K$, are known. Then, by plugging these data association solutions of all the $M$ BSs into problem (\ref{eqS5.5}), we can get a better estimation of the $K$ target locations, which is denoted by $(x_k^{(M)},y_k^{(M)})$, $k=1,\ldots,K$. According to (\ref{eqS5.4}), the overall cost for choosing the data association solutions of BSs 1, 2, and 3 as $\{\bar{g}_{1,k},\bar{g}_{2,k},\bar{g}_{3,k}\}_{k=1}^K \in \mathcal{G}^{(3)}$ is defined as
\begin{align}
&\Gamma(\{\bar{g}_{1,k},\bar{g}_{2,k},\bar{g}_{3,k}\}_{k=1}^K) \nonumber \\ = & \sum_{m=1}^M \sum_{k=1}^K \frac{(\mathcal{D}_m(\bar{g}_{m,k})-\sqrt{(a_m-x_k^{(M)})^2+(b_m-y_k^{(M)})^2})^2}{\sigma_{m,k}^2}.
 \label{eqS5.16}
\end{align}

At last, after searching all the feasible data association solutions of BSs 1, 2, and 3 in the set $\mathcal{G}_K^{(3)}$, we can select the one that can minimize the above overall error as follows
\begin{align}\label{eqn:optimal}
& \{g_{1,k}^\ast,g_{2,k}^\ast,g_{3,k}^\ast\}_{k=1}^K  \nonumber \\ = & \arg \min\limits_{\{g_{1,k},g_{2,k},g_{3,k}\}_{k=1}^K\in \mathcal{G}^{(3)}} \Gamma(\{g_{1,k},g_{2,k},g_{3,k}\}_{k=1}^K).
\end{align}
Then, the optimal data association solution of the other BSs and the optimal location solution of all the targets can be obtained via solving problem (\ref{eqS5.15}) and problem (\ref{eqS5.5}), respectively.

\begin{algorithm}[t]
{\bf Input}: $(a_m,b_m)$'s and $\mathcal{D}_m$'s, $\forall m \in \mathcal{M}$. \\
{\bf Initialization}: Obtain the feasible data association solutions for BSs 1, 2, and 3 in $\mathcal{G}^{(3)}$ given in (\ref{eqS5.8}). Define $\mathcal{G}^{(3)}(t)$ as the $t$-th data association solution in $\mathcal{G}^{(3)}$. Set $t=1$. \\
{\bf Repeat}:
\begin{enumerate}
\item[1.] Set $\mathcal{G}^{(3)}(t)$ as the data association solutions of BSs 1, 2, and 3, denoted by $\{\bar{g}_{1,k},\bar{g}_{2,k},\bar{g}_{3,k}\}_{k=1}^K$;
\item[2.] Solve problem (\ref{eqS5.5}) via the Gauss-Newton algorithm by setting $M=3$ to get an estimation of target locations via BSs 1, 2, and 3, denoted by $\{(x_k^{(3)},y_k^{(3)})\}_{k=1}^K$;
\item[3.] Calculate $\Delta d_{m,k,\tilde{k}}$'s based on (\ref{eqS5.11}), $\forall k,\tilde{k},m>3$, and solve problem (\ref{eqS5.15}) via the Hungarian algorithm to obtain the data association solution of BSs $4,\ldots,M$, denoted by $\{\bar{g}_{4,k},\ldots,\bar{g}_{M,k}\}_{k=1}^K$;
\item[4.] Solve problem (\ref{eqS5.5}) given $\{\bar{g}_{1,k},\ldots,\bar{g}_{M,k}\}_{k=1}^K$ to get a better estimation of target locations via all the BSs, denoted by $\{(x_k^{(M)},y_k^{(M)})\}_{k=1}^K$;
\item[5.] Calculate $\Gamma(\{\bar{g}_{1,k},\ldots,\bar{g}_{3,k}\}_{k=1}^K)$ based on (\ref{eqS5.16});
\item[6.] Set $t=t+1$.
\end{enumerate}
{\bf Until} $t=|\mathcal{G}^{(3)}|$. \\
{\bf Output}:

1). Obtain the optimal data association solutions for BSs 1, 2, and 3 via solving problem (\ref{eqn:optimal}), denoted by $\{g_{1,k}^\ast,g_{2,k}^\ast,g_{3,k}^\ast\}_{k=1}^K$.

2). Obtain the optimal data association solutions for BSs $4,\ldots,M$ via solving problem (\ref{eqS5.15}), denoted by $\{g_{4,k}^\ast,\ldots,g_{M,k}^\ast\}_{k=1}^K$.

3). Obtain the optimal locations of all the $K$ targets via solving problem (\ref{eqS5.5}), denoted by $\{(x_k^\ast,y_k^\ast)\}_{k=1}^K$.
\caption{ML-Based Algorithm to Solve Problem (\ref{eqS5.4}) for Target Localization}
\label{alg:2}
\end{algorithm}

The above procedure for solving problem (\ref{eqS5.4}) is summarized in Algorithm \ref{alg:2}. As compared to the exhaustive search based method, the complexity of Algorithm \ref{alg:2} is significantly reduced. First, instead of searching over all the $(K!)^{(M-1)}$ data association solutions of all the BSs that satisfy (\ref{eqn:BS 1}) and (\ref{eqn:matching}), under our proposed algorithm, we merely search over the feasible data association solutions of BSs 1, 2, and 3, i.e., $\mathcal{G}^{(3)}$ given in (\ref{eqS5.8}), as shown in problem (\ref{eqn:optimal}). Note that there are at most $(K!)^2$ solutions in $\mathcal{G}^{(3)}$ satisfying constraints (\ref{eqn:BS 1}) and (\ref{eqn:matching}); moreover, under constraints (\ref{eqS5.6}) and (\ref{eqS5.7}), the number of data association solutions in $\mathcal{G}^{(3)}$ is usually much smaller than $(K!)^2$. Second, under our proposed algorithm, given any data association solutions for BSs 1, 2, and 3, each BS $m>3$ can independently obtain its own data association solution by solving problem (\ref{eqS5.15}), instead of collaborating with the other BSs to jointly obtain their data association solutions.

\begin{remark}
If the target is detected by only part of BSs, the Algorithm \label{alg:2} is slightly modified. First, select three BSs to localize $\tilde{K} \leq K$ targets, denoted by $\{(x_{k}^{(3)},y_{k}^{(3)})\}_{k=1}^{\tilde{K}}$. Then for each BS $m$, apply Hungarian Algorithm to allocate $d_{m,k}$ to these $\tilde{K}$ targets and get indicator functions $\beta_{m,k,\tilde{k}}$'s. If the cost $\Delta d_{m,k,\tilde{k}}$ is larger than the threshold, then the corresponding $\beta_{m,k,\tilde{k}}$ is set to be zero. Next, based on the obtained $\beta_{m,k,\tilde{k}}$'s to estimate the $\tilde{K}$ targets. Last, eliminate these allocated distances from the distance sets $\mathcal{D}_m$ and repeat the above steps until all targets are localized.
\label{Remark 1}
\end{remark}

\subsection{Numerical Examples}
\begin{figure}
\centering
\subfigure[$r=1.5$ m]
{\scalebox{0.55}{\includegraphics*{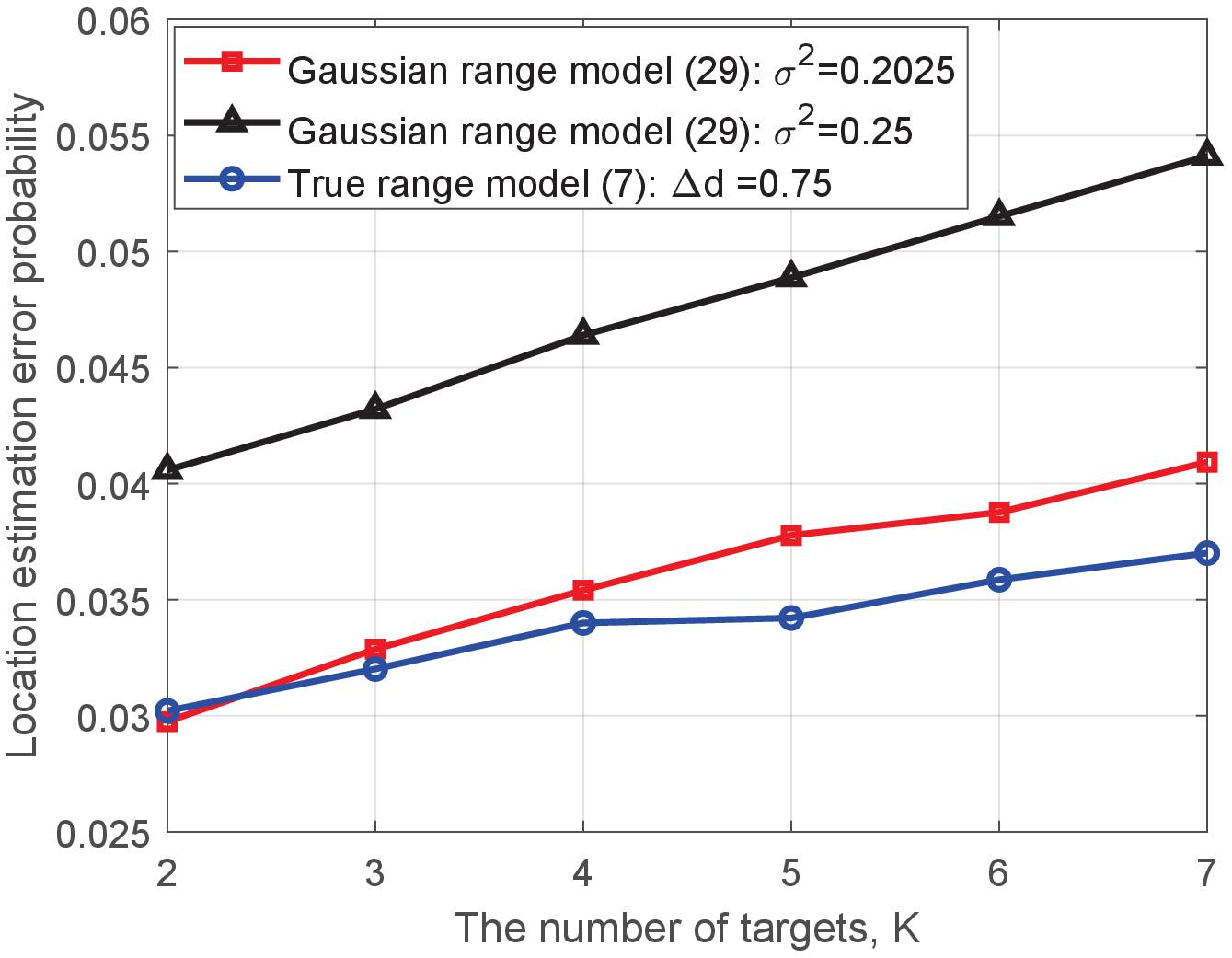}}}
\subfigure[$r=2.5$ m]
{\scalebox{0.55}{\includegraphics*{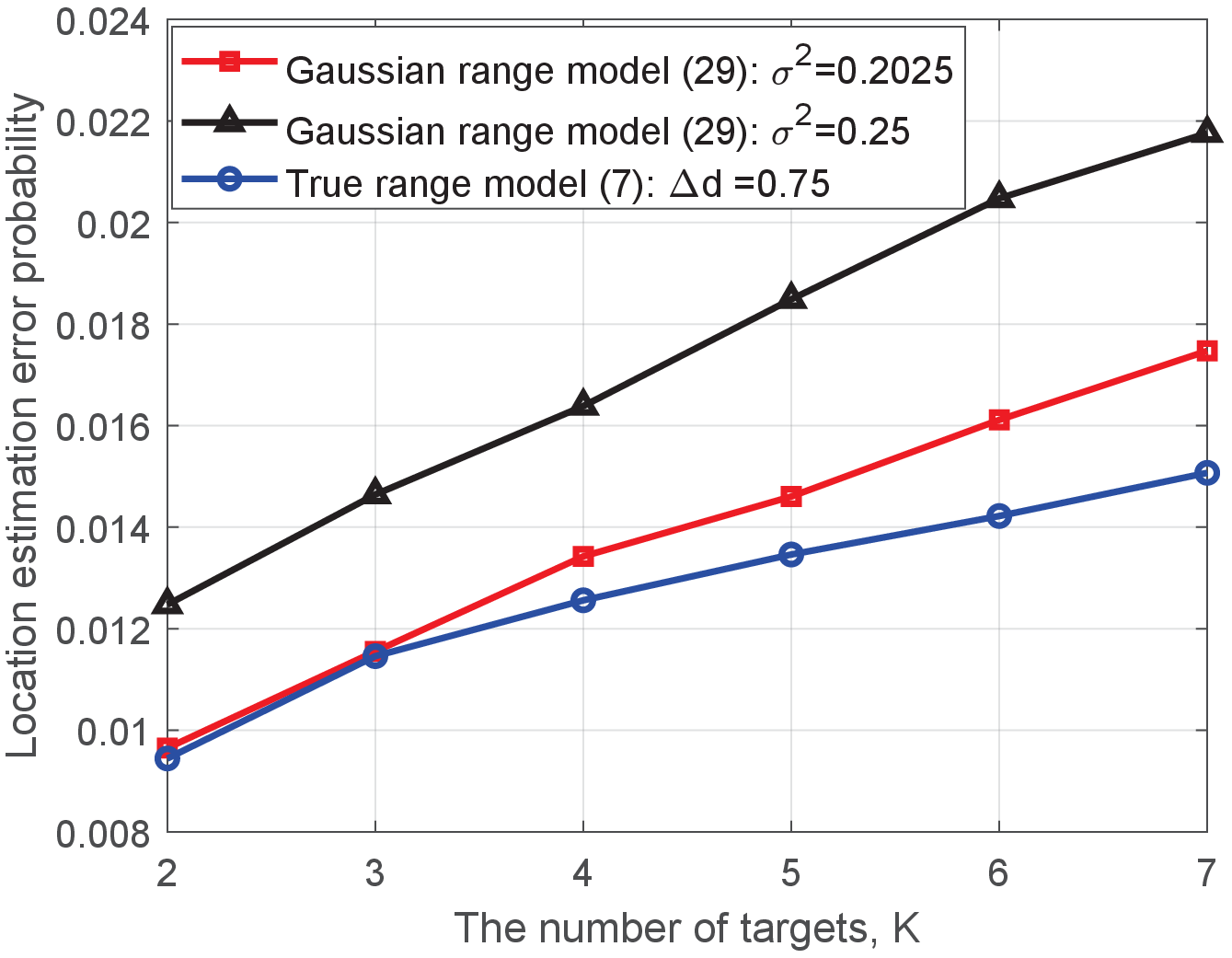}}}
\caption{Location estimation error probability versus the number of targets with $B=100$ MHz.}\label{fig7}
\end{figure}

In the following, we provide numerical examples to verify the effectiveness of Algorithm \ref{alg:2} for target localization, under the setup with $M=4$ BSs and $2\leq K\leq 7$ targets.

First, we consider the case where the channel bandwidth is $B=100$ MHz at the sub-6G band \cite{3gpp}, and the worst-case range estimation error shown in (\ref{eqS2.1}) is $\Delta d=0.75$ m. We assume that the BSs and targets are uniformly and randomly located in a $240$ m $\times$ $240$ m square, and generate $10^5$ independent realizations of their locations. In each realization, we first estimate $\mathcal{D}_m$'s based on (\ref{eqS2.10}), and then localize the targets by Algorithm \ref{alg:2} given $\mathcal{D}_m$'s, where an error event for localizing a target is defined as the case that the estimated location is not lying within a radius of $r$ m from the true target location. Let $N_{{\rm error}}$ denote the total number of error events in these $10^5$ realizations. Then, the location estimation error probability is defined as $\frac{N_{{\rm error}}}{K\times 10^5}$. Note that Algorithm \ref{alg:2} is designed based on the range estimation model in (\ref{eqS5.1}), where the estimation error is modeled as a Gaussian random variable, rather than the true range model in (\ref{eqS2.10}). To show that (\ref{eqS5.1}) is a good approximation of (\ref{eqS2.10}), in each realization, we also generate $\mathcal{D}_m$'s based on (\ref{eqS5.1}) with $\sigma_{m,k}^2=\sigma^2$, $\forall m,k$, as the input of Algorithm \ref{alg:2}, and evaluate the corresponding location estimation error probability.

Considering $r=1.5$ m and $r=2.5$ m, Fig. \ref{fig7} shows the location estimation error probability achieved by Algorithm \ref{alg:2} under the true range estimation model in (\ref{eqS2.10}) with worst-case error of $\Delta d=0.75$ m and the approximated model in (\ref{eqS5.1}) with $\sigma^2=0.2025$ or $\sigma^2=0.25$. It is observed that under the true range estimation model, the error probability to estimate the locations of $K=7$ targets is below $4\%$ and $1.6\%$ with $r=1.5$ m and $r=2.5$ m, respectively. Therefore, the estimation accuracy of our proposed scheme is in the order of meter with a probability higher than $95\%$ when the channel bandwidth is $B=100$ MHz. Moreover, it is observed that when $\sigma^2=0.2025$ in model (\ref{eqS5.1}), the performance achieved under this approximated model is very close to that achieved under the true range model (\ref{eqS2.10}). Thus, it is reasonable to use the Gaussian range model (\ref{eqS5.1}) in Algorithm \ref{alg:2} for localization in practical systems.

\begin{figure}
\centering
\subfigure[$r=0.6$ m]
{\scalebox{0.55}{\includegraphics*{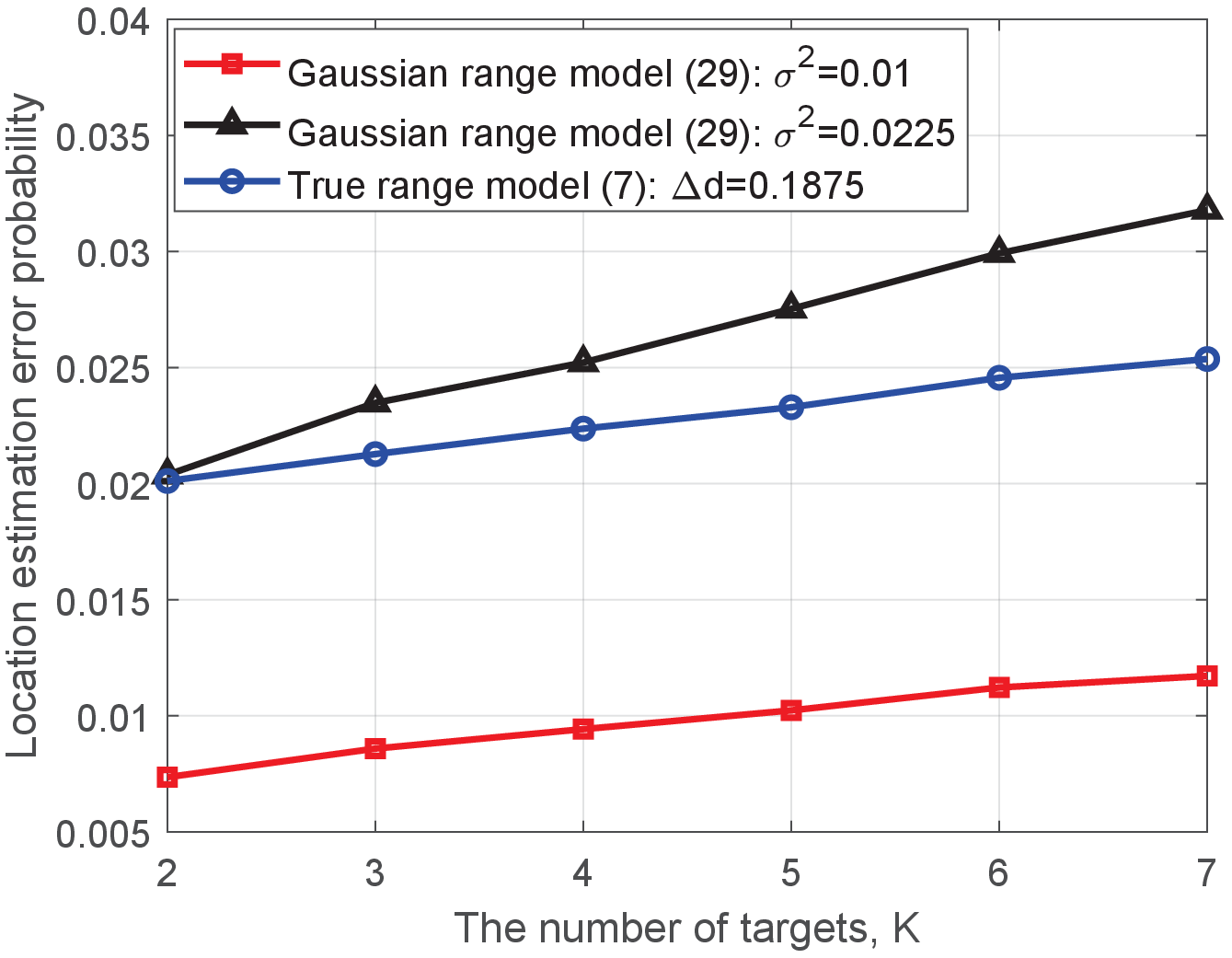}}}
\subfigure[$r=1$ m]
{\scalebox{0.55}{\includegraphics*{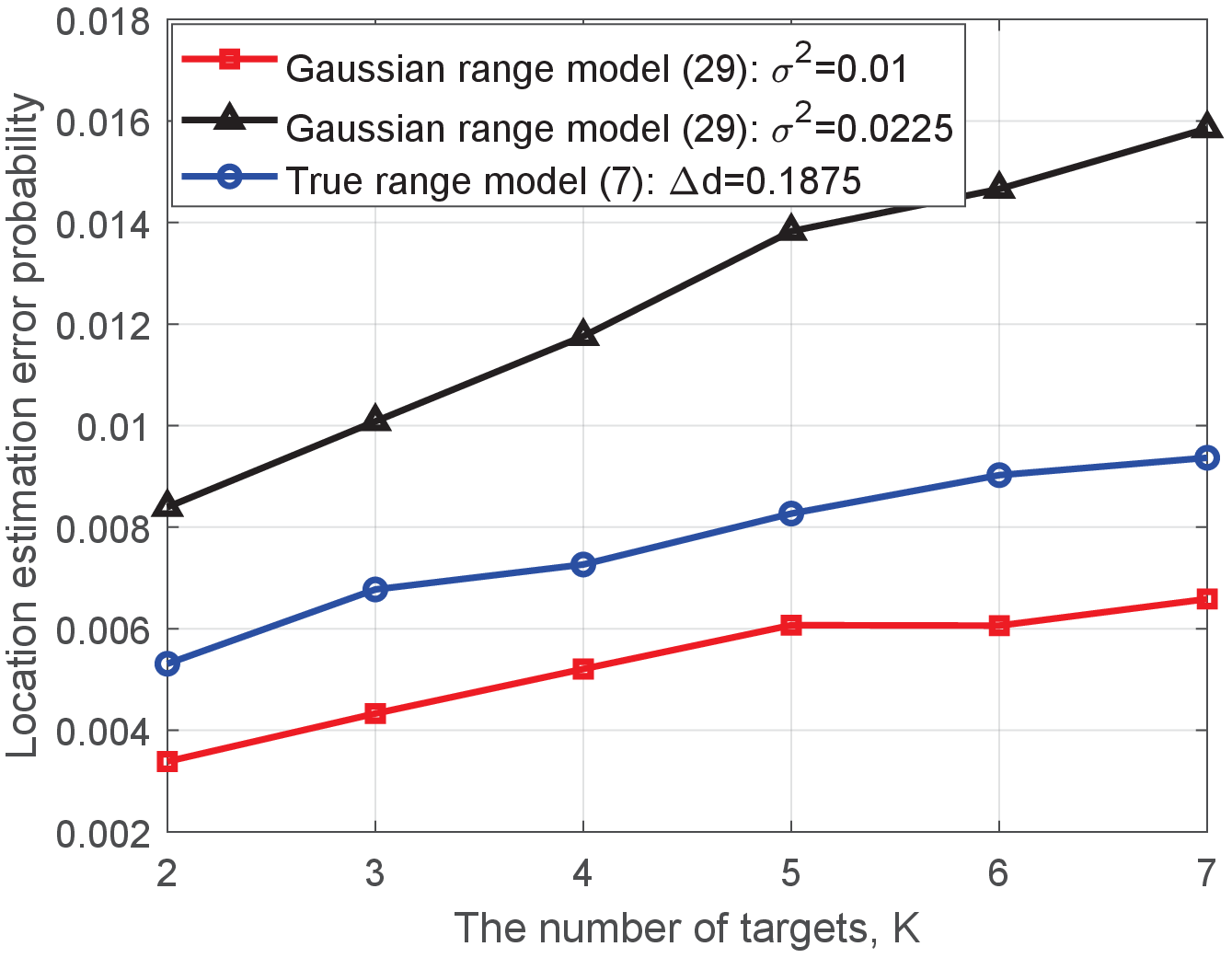}}}
\caption{Location estimation error probability versus the number of targets with $B=400$ MHz.}\label{fig9}
\end{figure}

Next, we consider the case where the channel bandwidth is $B=400$ MHz at the mmWave band \cite{3gpp}. In this case, the worst-case range estimation error shown in (\ref{eqS2.1}) is reduced to $\Delta d=0.1875$ m, and the length of CP is $0.585$ $\mu$s according to \cite{zaidi2017nr}. We assume that the BSs and users are located uniformly in an $80$ m $\times$ $80$ m square.\footnote{In this case, the maximum value of distance between a target and a BS such that the delay spread can be compensated by the CP (i.e., $L<Q$) is $0.585\times 10^{-6}\times c_0/2=87.75$ m, thus we consider an $80$ m $\times$ $80$ m region.} Moreover, we set $r=0.6$ m and $r=1$ m, respectively. Under the above setup, Fig. \ref{fig9} shows the location estimation error probability achieved by Algorithm \ref{alg:2} under the true range estimation model (\ref{eqS2.10}) with worst-case error of $\Delta d=0.1875$ m and the approximated model (\ref{eqS5.1}) with $\sigma^2=0.01$ or $\sigma^2=0.0225$. It is observed that the estimation accuracy of our proposed scheme is in the order of decimeter with a probability higher than $95\%$ when the channel bandwidth is $B=400$ MHz, thanks to the reduced worst-case estimation error due to the increased bandwidth.

\section{Concluding Remarks}\label{sec:Concluding Remarks}
In this paper, we proposed a novel two-phase framework for device-free sensing in an OFDM cellular network to achieve ISAC. Interesting theoretical results and practical algorithms were provided to deal with the data association issue to avoid the detection of ghost targets.

There are a number of directions along which the device-free sensing framework proposed in this paper can be further enriched. For example, it is interesting to investigate how to deploy the BSs in the cellular network not only to improve the communication throughput, but also to reduce the ghost target existence probability when the range estimation is imperfect. Moreover, in practice, signals reflected by other objects, e.g., buildings, may also be received by the sensing receive antennas. It is thus crucial to explore clutter suppression techniques to avoid such interference under the proposed framework.

\begin{appendix}

\subsection{Proof of Theorem \ref{Theorem2}}\label{appendix1}
Define $\mathcal{E}$ as the event that ghost target exists, i.e., there exists $\mathcal{X}^{{\rm G}}\neq \mathcal{X}$ satisfying (\ref{eqS4.6}) in Definition \ref{definition1}. Moreover, when event $\mathcal{E}$ occurs, define $\mathcal{E}_i$ as the event that $K-i$ true targets and $K-i$ ghost targets share the same coordinates, while the other $i$ ghost targets and $i$ true targets do not share any common coordinate, i.e., $|\mathcal{X}^{{\rm G}}\bigcap \mathcal{X}|=K-i$, $i=2,\ldots,K$.\footnote{If $K-1$ true targets and $K-1$ ghost targets share the same coordinates, the remaining true target and ghost target will also have the same coordinate since three BSs can uniquely locate one target. As a result, under event $\mathcal{E}$, $i=1$ will never happen.} We then have $\mathcal{E}=\bigcup_{i=2}^K \mathcal{E}_i$. Next, when event $\mathcal{E}_i$ occurs, based on which $i$ true targets possess different coordinates with the ghost targets, there are ${K\choose i}$ sub-events. Define $\mathcal{E}_i^{(r)}$ as the $r$-th sub-event, $r=1,\ldots,{K\choose i}$. Then, it follows that $\mathcal{E}_i=\bigcup_{r=1}^{{K\choose i}} \mathcal{E}_i^{(r)}$, $i=2,\ldots,K$. To summarize, we have
\begin{align}\label{eqn:probability}
{\rm Pr}(\mathcal{E})={\rm Pr}\left(\bigcup\limits_{i=2}^K\bigcup\limits_{r=1}^{{K\choose i}}\mathcal{E}_i^{(r)}\right)\leq \sum\limits_{i=2}^K\sum\limits_{r=1}^{{K\choose i}} {\rm Pr}(\mathcal{E}_i^{(r)}),
\end{align}where ${\rm Pr}(\mathcal{A})$ is the probability that event $\mathcal{A}$ happens.

\setcounter{equation}{46}
\begin{figure*}[ht]
	\begin{align}
		{\rm Pr}(\mathcal{E}_i^{(1)})&=  \int_{\{x_1,y_1,\ldots,x_i,y_i\}\in \mathcal{B}_i^{(1)}}p(x_1,y_1,\ldots,x_i,y_i) d x_1 dy_1\cdots d x_i d y_i \nonumber \\ &= \int_{\{x_1,y_1,\ldots,x_{i-1},y_{i-1}\}\in \tilde{\mathcal{B}}_i^{(1)}}\left(\int_{\{x_i,y_i\}\in \hat{\mathcal{B}}_i^{(1)}(x_1,y_1,\ldots,x_{i-1},y_{i-1})} p(x_i,y_i|x_1,y_1,\ldots,x_{i-1},y_{i-1})d x_i d y_i \right) \nonumber \\ & ~~~~~~~~~~~~~~~~~~~~~~~~~~~~~~~~~~~~~ p(x_1,y_1,\ldots,x_{i-1},y_{i-1}) d x_1 dy_1\cdots d x_{i-1} d y_{i-1} \label{eqn:pro 1} \\
		& =  \int_{\{x_1,y_1,\ldots,x_{i-1},y_{i-1}\}\in \tilde{\mathcal{B}}_i^{(1)}}\left(\int_{\{x_i,y_i\}\in \hat{\mathcal{B}}_i^{(1)}(x_1,y_1,\ldots,x_{i-1},y_{i-1})} p(x_i,y_i)d x_i d y_i \right) \nonumber \\ & ~~~~~~~~~~~~~~~~~~~~~~~~~~~~~~~~~~~~~ p(x_1,y_1,\ldots,x_{i-1},y_{i-1}) d x_1 dy_1\cdots d x_{i-1} d y_{i-1}, \label{eqn:pro 2}
	\end{align}
	\hrulefill
\end{figure*}

Without loss of generality, let us define sub-event $1$, i.e., $\mathcal{E}_i^{(1)}$, as the event that each of the coordinates of true targets $1,\ldots,i$ is not the coordinate of any ghost target, while each of the coordinates of true targets $i+1,\ldots,K$ is the coordinate of some ghost target. Note that error event $\mathcal{E}_i^{(1)}$ is equivalent to the event that in a system merely consisting of true target 1 to true target $i$ (without the other $K-i$ true targets), there exist $i$ ghost targets, whose coordinates are all different from the coordinates of the $i$ true targets. In the following, we show that ${\rm Pr}(\mathcal{E}_i^{(1)})=0$, $i=2,\ldots,K$. The similar approach can also be used to show that ${\rm Pr}(\mathcal{E}_i^{(r)})=0$, $\forall r\neq 1$.

Define $\mathcal{B}_i^{(1)}$ as the set of $(x_1,y_1,\ldots,x_i,y_i)$ such that if $(x_1,y_1,\ldots,x_i,y_i)\in \mathcal{B}_i^{(1)}$, the coordinates of the $i$ true targets with coordinates $(x_1,y_1),\ldots,(x_i,y_i)$ can lead to $i$ ghost targets with coordinates $(x_1^{{\rm G}},y_1^{{\rm G}}),\ldots,(x_i^{{\rm G}},y_i^{{\rm G}})$, where
\setcounter{equation}{45}
\begin{align}\label{co}
(x_q^{{\rm G}},y_q^{{\rm G}})\neq (x_k,y_k), ~ k,q=1,\ldots,i.
\end{align}Moreover, define $p(\cdot)$ as the probability density function (PDF). Then, we have (\ref{eqn:pro 1}) and (\ref{eqn:pro 2}) on the top of the next page, where $\tilde{\mathcal{B}}_i^{(1)}$ is the set of $(x_1,y_1,\ldots,x_{i-1},y_{i-1})$ such that if $(x_1,y_1,\ldots,x_{i-1},y_{i-1})\in \tilde{\mathcal{B}}_i^{(1)}$, there exists some $(x_i,y_i)$ to satisfy $(x_1,y_1,\ldots,x_{i},y_{i})\in \mathcal{B}_i^{(1)}$, and $\hat{\mathcal{B}}_i^{(1)}(x_1,y_1,\ldots,x_{i-1},y_{i-1})$ is the set of $(x_i,y_i)$ such that given any $(x_1,y_1,\ldots,x_{i-1},y_{i-1})\in \tilde{\mathcal{B}}_i^{(1)}$, if $(x_i,y_i)\in \hat{\mathcal{B}}_i^{(1)}(x_1,y_1,\ldots,x_{i-1},y_{i-1})$, then the coordinates of the $i$ true targets $(x_1,y_1),\ldots,(x_i,y_i)$ can lead to $i$ ghost targets with coordinates $(x_1^{{\rm G}},y_1^{{\rm G}}),\ldots,(x_i^{{\rm G}},y_i^{{\rm G}})$ satisfying (\ref{co}). In the above, (\ref{eqn:pro 2}) holds because all the targets are independently located in the network.

In the following, we prove that given any $\{x_1,y_1,\ldots,x_{i-1},y_{i-1}\}\in \tilde{\mathcal{B}}_i^{(1)}$, there are a finite number of elements in the set $\hat{\mathcal{B}}_i^{(1)}(x_1,y_1,\ldots,x_{i-1},y_{i-1})$. If this is true, then when $(x_i,y_i)$ is uniformly distributed in a continuous two-dimension region consisting of an infinite number of points, the probability that $(x_i,y_i)$ falls on the finite number of points in $\hat{\mathcal{B}}_i^{(1)}(x_1,y_1,\ldots,x_{i-1},y_{i-1})$ is zero, i.e., $\int_{\{x_i,y_i\}\in \hat{\mathcal{B}}_i^{(1)}(x_1,y_1,\ldots,x_{i-1},y_{i-1})} p(x_i,y_i)d x_i d y_i=0$, $\forall \{x_1,y_1,\ldots,x_{i-1},y_{i-1}\}\in \tilde{\mathcal{B}}_i^{(1)}$. This will indicate that ${\rm Pr}(\mathcal{E}_i^{(1)})=0$ according to (\ref{eqn:pro 2}) because $\int_{\{x_1,y_1,\ldots,x_{i-1},y_{i-1}\}\in \tilde{\mathcal{B}}_i^{(1)}} p(x_1,y_1,\ldots,x_{i-1},y_{i-1})$ $d x_1 dy_1\cdots d x_{i-1} d y_{i-1}$ is finite when target 1 to target $i-1$ are located uniformly in the network.

Define
\setcounter{equation}{48}
\begin{align}
\mathcal{D}_m^{(i)}=\{d_{m,1},\ldots,d_{m,i}\}, ~~~ \forall m,
\end{align}as the set consisting of the values of distance between BS $m$ and target 1 to target $i$, and $\mathcal{D}_m^{(i)}(g)$ as the $g$-th largest element in $\mathcal{D}_m^{(i)}$. Moreover, define $g_{m,k}^{(i)}$ such that $d_{m,k}=\mathcal{D}_m^{(i)}(g_{m,k}^{(i)})$, $k=1,\ldots,i$, $m=1,\ldots,M$. Given any $\{x_1,y_1,\ldots,x_{i-1},y_{i-1}\}\in \tilde{\mathcal{B}}_i^{(1)}$, $i-1$ elements in $\mathcal{D}_m^{(i)}$ are fixed, $\forall m$. Define the range set to localize target $i$ as
\begin{align}\label{eqn:target}
\mathcal{T}_i=\{\mathcal{D}_1^{(i)}(g_{1,i}^{(i)}),\ldots,\mathcal{D}_M^{(i)}(g_{M,i}^{(i)})\},
\end{align}which consists of the remaining one variable range of each BS. In the following, we study all the possibilities of $\mathcal{T}_i$ such that with $(x_i,y_i)$ localized by $\mathcal{T}_i$ and the given $\{x_1,y_1,\ldots,x_{i-1},y_{i-1}\}\in \tilde{\mathcal{B}}_i^{(1)}$, $i$ ghost targets with coordinates $(x_1^{{\rm G}},y_1^{{\rm G}}),\ldots,(x_i^{{\rm G}},y_i^{{\rm G}})$ satisfying (\ref{co}) exist.

Consider a data association solution $\{\bar{g}_{m,1}^{(i)},\ldots,\bar{g}_{m,i}^{(i)}\}_{m=1}^M$ that is different from the correct data association solution $\{g_{m,1}^{(i)},\ldots,g_{m,i}^{(i)}\}_{m=1}^M$ and satisfies $\{\bar{g}_{m,1}^{(i)},\ldots,\bar{g}_{m,i}^{(i)}\}=\{1,\ldots,i\}$, $\forall m$. Note that there are $(i!)^M-1$ data association solutions satisfying the above conditions. According to Definition \ref{definition1}, a data association solution can lead to $i$ ghost targets if and only if for each $q\in \{1,\ldots,i\}$, there exists a solution $(x_q^{{\rm G}},y_q^{{\rm G}})$ to the following $M$ equations:
\begin{align}\label{eqn:loc}
\sqrt{(a_m-x_q^{{\rm G}})^2+(b_m-y_q^{{\rm G}})^2}=\mathcal{D}_m^{(i)}(\bar{g}_{m,q}^{(i)}), ~ \forall m\in \mathcal{M}.
\end{align}Note that if we can find a $q\in \{1,\ldots,i\}$ such that $\mathcal{D}_m^{(i)}(\bar{g}_{m,q}^{(i)})=\mathcal{D}_m^{(i)}(g_{m,i}^{(i)})$ holds for at least three values of $m$, then it indicates that $(x_q^{{\rm G}},y_q^{{\rm G}})=(x_i,y_i)$, because true target $i$ and ghost target $q$ have the same distance values to three BSs, and three BSs not on the same line can localize a unique target. This violates the definition of $\mathcal{B}_i^{(1)}$ where all true targets and ghost targets have different coordinates, i.e., (\ref{co}). As a result, for any $q\in \{1,\ldots,i\}$, $\mathcal{D}_m^{(i)}(\bar{g}_{m,q}^{(i)})=\mathcal{D}_m^{(i)}(g_{m,i}^{(i)})$ holds for at most two values of $m$, denoted by $m_{q,1}$ and/or $m_{q,2}$, and if $m\in \mathcal{M}_q=\{m|m\in \mathcal{M}, m\neq m_{q,1},m\neq m_{q,2}\}$, we have $\mathcal{D}_m^{(i)}(\bar{g}_{m,q}^{(i)})=\mathcal{D}_m^{(i)}(g_{m,k}^{(i)})$ for some $k\in \{1,\ldots,i-1\}$, where $\mathcal{D}_m^{(i)}(g_{m,k}^{(i)})$'s are known given $\{x_1,y_1,\ldots,x_{i-1},y_{i-1}\}\in \tilde{\mathcal{B}}_i^{(1)}$. Note that if $M \geq 4$, $|\mathcal{M}_q|\geq 2$, $\forall q$. As a result, for any $q$, the following $|\mathcal{M}_q|\geq 2$ equations with known $\mathcal{D}_m^{(i)}(\bar{g}_{m,q}^{(i)})$'s have at most two possible solutions for the coordinate of ghost target $q$:
\begin{align}
\sqrt{(a_m-x_q^{{\rm G}})^2+(b_m-y_q^{{\rm G}})^2}=\mathcal{D}_m^{(i)}(\bar{g}_{m,q}^{(i)}), ~ \forall m\in \mathcal{M}_q.
\end{align}Then, for any $q$, there are at most two possible values of $\mathcal{D}_{m_{q,1}}^{(i)}(g_{m_{q,1},i}^{(i)})=\mathcal{D}_{m_{q,1}}^{(i)}(\bar{g}_{m_{q,1},q}^{(i)})$ and at most two possible values of $\mathcal{D}_{m_{q,2}}^{(i)}(g_{m_{q,2},i}^{(i)})=\mathcal{D}_{m_{q,2}}^{(i)}(\bar{g}_{m_{q,2},q}^{(i)})$. Note that $\bigcup_{q=1}^i \{m_{q,1},m_{q,2}\}=\mathcal{M}$, since if ghost target exists, for each $m$, there must exist some $q$ such that $\mathcal{D}_m^{(i)}(\bar{g}_{m,q}^{(i)})=\mathcal{D}_m^{(i)}(g_{m,i}^{(i)})$ according to Definition \ref{definition1}. As a result, given each data association solution $\{\bar{g}_{m,1}^{(i)},\ldots,\bar{g}_{m,i}^{(i)}\}_{m=1}^M$, $\mathcal{D}_m^{(i)}(g_{m,i}^{(i)})$ has at most two possible values, $\forall m$, and there are thus at most $2^M$ possibilities for the set $\mathcal{T}_i$ defined in (\ref{eqn:target}). Recall that there are $(i!)^M-1$ feasible data association solutions of $\{\bar{g}_{m,1}^{(i)},\ldots,\bar{g}_{m,i}^{(i)}\}_{m=1}^M$. Therefore, given any $\{x_1,y_1,\ldots,x_{i-1},y_{i-1}\}\in \tilde{\mathcal{B}}_i^{(1)}$, we have
\begin{align}
|\hat{\mathcal{B}}_i^{(1)}(x_1,y_1,\ldots,x_{i-1},y_{i-1})|\leq ((i!)^M-1)2^M.
\end{align}

To summarize, when $M\geq 4$, given any $\{x_1,y_1,\ldots,x_{i-1},y_{i-1}\}\in \tilde{\mathcal{B}}_i^{(1)}$, the number of elements in the set $\hat{\mathcal{B}}_i^{(1)}(x_1,y_1,\ldots,x_{i-1},y_{i-1})$ is finite. As stated in the above, this indicates that ${\rm Pr}(\mathcal{E}_i^{(1)})=0$ in (\ref{eqn:pro 2}). Similarly, we can prove that ${\rm Pr}(\mathcal{E}_i^{(r)})=0$, $\forall r\neq 1$. According to (\ref{eqn:probability}), it follows that ${\rm Pr}(\mathcal{E})=0$. Theorem \ref{Theorem2} is thus proved.

\subsection{Proof of Lemma \ref{Lemma1}}\label{appendix2}
First, we show that two necessary conditions for the existence of ghost targets are as follows
\begin{align}
&\bigcup_{q \in \mathcal{K}} \mathcal{S}_{k,q} = \mathcal{M},\quad \forall k \in \mathcal{K},\label{eqS4.11}
\\
&\bigcup_{k \in \mathcal{K}} \mathcal{S}_{k,q} = \mathcal{M},\quad \forall q \in \mathcal{K}. \label{eqS4.12}
\end{align}Specifically, given some $\mathcal{X}^{{\rm G}}\neq \mathcal{X}$, suppose that (\ref{eqS4.11}) does not hold. In this case, suppose that there exist $\bar{m}\in \mathcal{M}$ and $\bar{k}\in \mathcal{K}$ such that $\bar{m}\notin \bigcup_{q\in \mathcal{K}}\mathcal{S}_{\bar{k},q}$. This indicates that $d_{\bar{m},\bar{k}}$ is not in the set $\mathcal{D}_{\bar{m}}^{{\rm G}}$. In other words, $\mathcal{D}_{\bar{m}}\neq \mathcal{D}_{\bar{m}}^{{\rm G}}$. As a result, for any $\mathcal{X}^{{\rm G}}\neq \mathcal{X}$ such that (\ref{eqS4.11}) does not hold, it cannot be the set of coordinates of ghost targets according to Definition \ref{definition1}. Similarly, for any $\mathcal{X}^{{\rm G}}\neq \mathcal{X}$ such that (\ref{eqS4.12}) does not hold, it cannot be the set of coordinates of ghost targets. To summarize, if $\mathcal{X}^{{\rm G}}\neq \mathcal{X}$ is the set of coordinates of ghost targets, then (\ref{eqS4.11}) and (\ref{eqS4.12}) should hold.

In the following, we prove Lemma \ref{Lemma1} with $M=4$ and $K=2$ based on the above two necessary conditions. First, consider the case when $\mathcal{X}^{{\rm G}}\neq \mathcal{X}$ satisfies that $\mathcal{X}^{{\rm C}}$ defined in (\ref{eqn:common set}) is not an empty set, i.e., $(x_k,y_k)=(x_q^{{\rm G}},y_q^{{\rm G}})$ for some $k,q\in \mathcal{K}$. We show by contradictory that in this case, $\mathcal{X}^{{\rm G}}$ cannot be the coordinates of ghost targets. Suppose that $\mathcal{X}^{{\rm G}}$ consists of the coordinates of ghost targets. Then, conditions (\ref{eqS4.11}) and (\ref{eqS4.12}) indicate that $d_{m,\bar{k}}=d_{m,\bar{q}}$, $m=1,2,3,4$, where $\bar{k}\in \mathcal{K}\neq k$ and $\bar{q}\in \mathcal{K}\neq q$. Since any 3 BSs are not on the same line, $d_{m,\bar{k}}=d_{m,\bar{q}}$, $m=1,2,3,4$, indicates that $(x_{\bar{k}},y_{\bar{k}})= (x_{\bar{q}}^{{\rm G}},y_{\bar{q}}^{{\rm G}})$. Together with $(x_k,y_k)=(x_q^{{\rm G}},y_q^{{\rm G}})$, this contradicts to the fact that $\mathcal{X}^{{\rm G}}\neq \mathcal{X}$. As a result, if $\mathcal{X}^{{\rm G}}\neq \mathcal{X}$ satisfies that $\mathcal{S}_{k,q}=\mathcal{M}$ for some $k$ and $q$, then $\mathcal{X}^{{\rm G}}$ cannot be the coordinates of ghost targets.

Next, consider the case when $\mathcal{X}^{{\rm G}}\neq \mathcal{X}$ satisfies that $\mathcal{X}^{{\rm C}}$ defined in (\ref{eqn:common set}) is an empty set. In the following, we show the necessary conditions for the existence of ghost targets in this case. Suppose that $\mathcal{X}^{{\rm G}}$ consists of the coordinates of ghost targets. Then, conditions (\ref{eqS4.11}) and (\ref{eqS4.12}) indicate that
\begin{align}
&  \sum_{q=1}^2 |\mathcal{S}_{k,q}|=4, \quad \forall k\in \mathcal{K}, \label{eqC3.1} \\
&  \sum_{k=1}^2 |\mathcal{S}_{k,q}|=4. \quad \forall q\in \mathcal{K}. \label{eqC3.2}
\end{align}
Because any 3 BSs are not on the same line and $\mathcal{X}^{{\rm C}}$ defined in (\ref{eqn:common set}) is an empty set, we have $|\mathcal{S}_{k,q}|\leq 2$, $\forall k,q$. To satisfy (\ref{eqC3.1}) and (\ref{eqC3.2}), we must have $|\mathcal{S}_{k,q}| = 2$, $\mathcal{S}_{k,1}\bigcap \mathcal{S}_{k,2}=\emptyset$, and $\mathcal{S}_{1,q}\bigcap \mathcal{S}_{2,q}=\emptyset$, $\forall k, q$. Then, it follows that $\mathcal{S}_{1,1}=\mathcal{S}_{2,2}=\{1,2,3,4\}/\mathcal{S}_{1,2}$ and $\mathcal{S}_{1,2}=\mathcal{S}_{2,1}=\{1,2,3,4\}/\mathcal{S}_{1,1}$. $\mathcal{S}_{1,1}=\mathcal{S}_{2,2}$ and $|\mathcal{S}_{1,1}|=|\mathcal{S}_{2,2}|=2$ require that the line connecting the two BSs in $\mathcal{S}_{1,1}=\mathcal{S}_{2,2}$ is the perpendicular bisector of the line segment connecting $(x_1,y_1)$ and $(x_1^{{\rm G}},y_1^{{\rm G}})$ as well as the line segment connecting $(x_2,y_2)$ and $(x_2^{{\rm G}},y_2^{{\rm G}})$. Similarly, we can show based on $\mathcal{S}_{1,2}=\mathcal{S}_{2,1}$ and $|\mathcal{S}_{1,2}|=|\mathcal{S}_{2,1}|=2$ that the line connecting the two BSs in $\mathcal{S}_{1,2}=\mathcal{S}_{2,1}$ is the perpendicular bisector of the line segment connecting $(x_1,y_1)$ and $(x_2^{{\rm G}},y_2^{{\rm G}})$ as well as the line segment connecting $(x_2,y_2)$ and $(x_1^{{\rm G}},y_1^{{\rm G}})$. The above shows that the necessary conditions for the existence of the ghost targets are as follows: 1. the line connecting the BSs in $\mathcal{S}_{1,1}=\mathcal{S}_{2,2}$ is perpendicular to that connecting the BSs in $\mathcal{S}_{1,2}=\mathcal{S}_{2,1}$; and 2. the line segment connecting $(x_1,y_1)$ and $(x_1^{{\rm G}},y_1^{{\rm G}})$, that connecting $(x_2,y_2)$ and $(x_2^{{\rm G}},y_2^{{\rm G}})$, that connecting $(x_1,y_1)$ and $(x_2^{{\rm G}},y_2^{{\rm G}})$, and that connecting $(x_2,y_2)$ and $(x_1^{{\rm G}},y_1^{{\rm G}})$ form a rectangle. As a result, if the first necessary condition does not hold, there never exist the ghost targets. On the other hand, if the first necessary condition holds, we show that the probability that the second necessary condition holds is zero when the two targets are located uniformly in the network. Let $(x_0,y_0)$ denote the intersection point of the two perpendicular lines that connect the BSs in $\mathcal{S}_{1,1}=\mathcal{S}_{2,2}$ and connect the BSs in $\mathcal{S}_{1,2}=\mathcal{S}_{2,1}$. If the second necessary condition is true, then we have $x_1+x_2=2x_0$ and $y_1+y_2=2y_0$, which define a two-dimension plane in the four-dimension space for $x_1,x_2,y_2,y_2$. If the two targets are located uniformly in the network, $x_1+x_2=2x_0$ and $y_1+y_2=2y_0$ occur with probability zero. As a result, if the first necessary condition is true, there exist no ghost targets almost surely if the targets are located uniformly in the network.

By combining the cases when $\mathcal{X}^{{\rm C}}$ is not an empty set and is an empty set, Lemma \ref{Lemma1} is thus proved.

\end{appendix}

\bibliographystyle{IEEEtran}
\bibliography{ISAC2022}

\end{document}